\documentclass[pre,12pt,onecolumn,tightenlines]{revtex4}  % for review and submission
\usepackage{amsfonts}
\usepackage{amsmath}
\usepackage{amssymb}
\usepackage{version}
\usepackage{graphicx}%
\includeversion{New_connection}
\excludeversion{Old_connection}

\begin{document}
%\today
%\preprint{HEP/123-qed}
\title{Complete Set of Stochastic Verlet-Type Thermostats for Correct Langevin Simulations}

\author{$\;$ \\ Niels Gr{\o}nbech-Jensen}
\email{ngjensen@math.ucdavis.edu}
\affiliation{Department of Mechanical \& Aerospace Engineering --- Department of Mathematics\\ University of California, Davis, CA 95616, U.S.A.}
%\keywords{Molecular Dynamics, Discrete-Time Langevin Equations, St{\o}rmer-Verlet Algorithms, Computational Statistical Mechanics, Stochastic Thermostats}
%\today

$\;$\\

\begin{abstract}

\vspace{0.3 in}

\begin{center}
{\large Abstract}
\end{center}

\vspace{0.1 in}

\noindent
We present the complete set of stochastic Verlet-type algorithms that can provide correct statistical measures for both configurational and kinetic sampling in discrete-time Langevin systems. The approach is a brute-force general representation of the Verlet-algorithm with free parameter coefficients that are determined by requiring correct Boltzmann sampling for linear systems, regardless of time step. The result is a set of statistically correct methods given by one free functional parameter, which can be interpreted as the one-time-step velocity attenuation factor.  We define the statistical characteristics of both true on-site $v^n$ and true half-step $u^{n+\frac{1}{2}}$ velocities, and use these definitions for each statistically correct St{\o}rmer-Verlet method to find a {\it unique} associated half-step velocity expression, which yields correct kinetic Maxwell-Boltzmann statistics for linear systems. It is shown that no other similar, statistically correct on-site velocity exists. We further discuss the use and features of finite-difference velocity definitions that are neither true on-site, nor true half-step. The set of methods is written in convenient and conventional stochastic Verlet forms that lend themselves to direct implementation for, e.g., Molecular Dynamics applications. We highlight a few specific examples, and validate the algorithms through comprehensive Langevin simulations of both simple nonlinear oscillators and complex Molecular Dynamics.
\end{abstract}
\maketitle
\vspace{1.5 in}
\begin{center}
{\small{To Appear in {\it Molecular Physics}:  https://doi.org/10.1080/00268976.2019.1662506}}
\end{center}
\newpage

\section{Introduction}
\label{sec:intro}
Over the past several decades, discrete-time Langevin and Molecular Dynamics (MD) simulations have provided a wealth of information about the properties of nonlinear and complex systems \cite{AllenTildesley,Frenkel,Rapaport,Hoover_book,Leach}. While the simulations are intended to represent the continuous-time equations of motion, the inevitable temporal discretization alters not only the accuracy of the simulated trajectories, but in some cases also fundamental aspects of the system itself. Thus, an integral part of any simulation task is to explore and optimize the balance between the two conflicting objectives; namely simulation efficiency by increasing the discrete time step, and simulation accuracy by decreasing the discrete time step. For computational statistical mechanics, one of the key equations of motion to investigate is the Langevin equation \cite{Langevin}, which is the topic we are concerned with in this paper,
\begin{eqnarray}
m\dot{v}+\alpha\dot{r} & = & f+\beta \; . \label{eq:Langevin}
\end{eqnarray}
Here $m$ is the mass of an object with spatial (configurational) coordinate $r$ and velocity $v=\dot{r}$. The object is subjected to a force $f$ and linear friction, which is represented by the non-negative constant $\alpha$. The fluctuation-dissipation relationship specifies that the thermal fluctuations $\beta$ can be represented by the Gaussian distribution \cite{Parisi}
\begin{subequations}
\begin{eqnarray}
\langle\beta(t)\rangle & = & 0 \\
\langle\beta(t)\beta(t^\prime)\rangle & = & 2\alpha k_BT \delta(t-t^\prime) \; , 
\end{eqnarray}
\end{subequations}
where $\delta(t)$ is Dirac's delta function, $k_B$ is Boltzmann's constant, and $T$ is the thermodynamic temperature.

Many methods for controlling the temperature of a simulated system (ÒthermostatsÓ) have been developed, and most of them fall into two major categories: Deterministic (e.g., Nos{\'e}-Hoover \cite{Nose,Hoover,Hoover2,Martyna_92,Watanabe}) and stochastic (Langevin) thermostats (see the large body of work represented in, e.g., Refs.~\cite{SS,vgb_1982,Gunsteren,BBK,Skeel_2002,rc_2003,Melchionna_2007,Bussi_2007,Bussi_07,Loncharich_2004,thalmann_2007,Mishra_1996,vdSpoel_2005,Vanden,Goga_2012,Paquet}). The deterministic approach includes additional degrees of freedom, which act as an energy reservoir and thereby mimic a thermal heat bath. A requirement for such method is that the temperature of a simulated system can be reliably measured in order for the method to interact properly with the heat-bath. The stochastic approach is to directly simulate the Langevin equation, Eq.~(\ref{eq:Langevin}), which does not include additional degrees of freedom, but instead interacts with a heat-bath through direct inclusion of noise given by the fluctuation-dissipation balance for $\alpha>0$. The methods commonly display time step errors of order one or two, either in configurational or kinetic sampling results, or both \cite {Pastor_88,skeel_2003}. The work described in this paper exclusively considers the stochastic Langevin approach within the Verlet framework outlined below.

Accurate numerical solutions to the Langevin equation can be found by a variety of discrete-time algorithms for small time steps, since this limit will approach the correct continuous-time equation. As the time step is increased, these solutions will increasingly deviate from physically meaningful trajectories. However, in computational statistical mechanics it is often not essential to attain precise trajectories of the simulated objects. Rather, the core objective is typically to obtain trajectories that sample a physically meaningful ensemble, which has the correct statistical properties. Therefore, while it is both necessary and desirable to make certain that a simulation becomes physically accurate in its details in the continuous-time limit, algorithm development for simulating Langevin equations are keenly focused on ensuring correct statistics in a broader range of applied time steps. This objective, however, is complicated by an inherent feature of discrete time; namely the absence of access to the true velocity that represents the simulated discrete-time trajectory (see, e.g.,  Refs.~\cite{Venneri,holian95,GJF3,2GJ}). As a result, the simulated configurational and kinetic statistics of a system are generally mutually inconsistent \cite{Pastor_88}. In the past decade, stochastic thermostats (i.e., discrete-time algorithms for the Langevin equation) have been constructed such that configurational statistics is correct for linear systems (see the GJF method Ref.~\cite{GJF1} and later Ref.~\cite{Sivak}, which made a temporal revision to Ref.~\cite{ML} in order to correct the diffusive component of configurational sampling). Common for those methods is that the kinetic statistics is incorrectly measured, and the error scales with the square of the time step. An important component to solving this problem is to realize that, in discrete-time, the evolution of the configurational coordinate $r$ can be represented without its associated velocity.

{\underline{For $\alpha=0$}} (the Newtonian limit) in Eq.~(\ref{eq:Langevin}) this can be illustrated by the second order discrete-time St{\o}rmer-Verlet approximation \cite{Stormer_1921,Verlet,Gear}, which is solely represented by the discrete-time configurational degree of freedom $r^n$
\begin{eqnarray}
r^{n+1} & = & 2r^n-r^{n-1}+\frac{dt^2}{m}f^n \; , \label{eq:sv_r_basic}
\end{eqnarray}
where the superscript $n$ of the variables $r^n$ and $f^n=f(t_n,r^n)$ represents the discrete-time $t_n=t_0+n\,dt$, $dt$ being the discrete time step.
Since this is a second order difference equation, all configurational properties of the numerical solution using this algorithm are given without the need for a velocity variable. Without changing the configurational trajectory $r^n$, given by Eq.~(\ref{eq:sv_r_basic}), different velocity variables can be attached to $r^n$ in order to approximate the associated velocity. For this simple Newtonian case, one obvious choice is the second order, finite difference {\it on-site} velocity \cite{AllenTildesley,Swope,Beeman} $v^n$
\begin{eqnarray}
v^n & = & \frac{r^{n+1}-r^{n-1}}{2dt} \; , \label{eq:sv_v_basic}
\end{eqnarray}
which, combined with Eq.~(\ref{eq:sv_r_basic}), can be expressed as the Velocity-explicit Verlet (VV) method
\begin{subequations}
\begin{eqnarray}
r^{n+1} & = & r^n+dt\,v^n+\frac{dt^2}{2m}f^n \label{eq:vv_r_basic}\\
v^{n+1} & = & v^n+\frac{dt}{2m}(f^n+f^{n+1}) \; . \label{eq:vv_v_basic}
\end{eqnarray}
\label{eq:vv_basic}
\end{subequations}
The other obvious second order, finite difference choice \cite{AllenTildesley,Buneman,Hockney} defines a {\it half-step} velocity $v^{n+\frac{1}{2}}$ by
\begin{eqnarray}
v^{n+\frac{1}{2}} & = & \frac{r^{n+1}-r^n}{dt}\; , \label{eq:sv_u_basic}
\end{eqnarray}
which, combined with Eq.~(\ref{eq:sv_r_basic}), can be expressed as the leap-frog (LF) method
\begin{subequations}
\begin{eqnarray}
v^{n+\frac{1}{2}} & = & v^{n-\frac{1}{2}}+\frac{dt}{m}f^n \label{eq:lf_v_basic}\\
r^{n+1} & = & r^n+dt\, v^{n+\frac{1}{2}} \; .\label{eq:lf_r_basic}
\end{eqnarray}
\label{eq:lf_basic}
\end{subequations}
While the configurational trajectory $r^n$ is seemingly interacting with a velocity variable in Eqs.~(\ref{eq:vv_basic}) and (\ref{eq:lf_basic}), the resulting values of $r^n$ found from Eqs.~(\ref{eq:sv_r_basic}), (\ref{eq:vv_r_basic}), and (\ref{eq:lf_r_basic}) are identical. The two velocity definitions shown above are different, not only in the time at which they approximate the continuous time velocity, but also in the quality of the approximation since they are central difference approximations over time intervals of different sizes.  Thus, the velocity variable is a non-unique approximation to a velocity of a trajectory in discrete time, and while a velocity depends on the trajectory, e.g., through Eq.~(\ref{eq:sv_v_basic}) or (\ref{eq:sv_u_basic}), the configurational trajectory $r^n$ does not depend on the definition of the associated velocity. In fact, one may simultaneously associate more than one velocity with a trajectory, as can be seen in the set of equations \cite{Tuckerman_92}
\begin{subequations}
\begin{eqnarray}
v^{n+\frac{1}{2}} & = & v^n +\frac{dt}{2m}f^n \label{eq:compact_u_basic}\\
r^{n+1} & = & r^n+dt\,v^{n+\frac{1}{2}}\label{eq:compact_r_basic}\\
v^{n+1} & = & v^{n+\frac{1}{2}}+\frac{dt}{2m}\,f^{n+1} \; , \label{eq:compact_v_basic}
\end{eqnarray}
\label{eq:compact_basic}
\end{subequations}
which is an identical alternate representation of Eqs.~(\ref{eq:sv_r_basic}), (\ref{eq:sv_v_basic}), and (\ref{eq:sv_u_basic}). This ambiguity of multiple choices in defining a velocity approximation in discrete time opens an opportunity to design velocities with certain discrete-time characteristics without altering the underlying configurational discrete-time trajectory.

Exploiting the ambiguity for the Langevin system, Eq.~(\ref{eq:Langevin}) with $\alpha>0$, it was recently shown \cite{2GJ} that a particular definition of an accompanying discrete-time velocity, the 2GJ (Gr{\o}nbech Jensen \& Gr{\o}nbech-Jensen) velocity, to the time-step-independent and configurationally correct GJF (Gr{\o}nbech-Jensen \& Farago) trajectory \cite{GJF1,GJF2} can be identified such that also kinetic statistics is correctly measured for linear systems for all time steps within the stability range. The remainder of this paper will use the term "correct" in reference to statistically correct results for linear systems. The result of Ref.~\cite{2GJ} is a practical and convenient discrete-time thermostat for Langevin systems expressed in the common form of the Verlet-type algorithms, and the applicability of this GJF-2GJ thermostat was demonstrated for both simple, nonlinear systems as well as for complex Molecular Dynamics. Based on this discovery, we recall that the work presented in Refs.~\cite{ML,Sivak} led to an algorithm, which also produced correct, time-step-independent configurational statistics for linear systems. Thus, given that there exist more than one statistically correct configurational trajectory, the obvious question is: How many different kinds of statistically correct methods exist, and how do we define statistically correct velocity variables associated with those trajectories? In this paper 1) we find all possibilities of finite-difference, stochastic St{\o}rmer-Verlet-type thermostats that yield correct configurational Boltzmann statistics and Einstein diffusion for linear systems, and 2) for each of those statistically correct St{\o}rmer-Verlet methods, we find all possible meaningful, associated velocity expressions that yield correct kinetic Maxwell-Boltzmann statistics. In conjunction with these results we also provide specific definitions of how to identify proper on-site and half-step velocities with meaningful statistical coordination to the underlying spatial St{\o}rmer-Verlet trajectory when the system has friction and noise.

The structure of the presentation is as follows: In Section~\ref{sec:review} we briefly review the features and expressions of the GJF and GJF-2GJ methods, since they are the methods to be generalized in this work. Based on this review, we then, in Sec.~\ref{sec:sv_section}, derive the complete set of St{\o}rmer-Verlet expressions for statistically correct trajectories without addressing the velocity component of the statistical properties. In that section we also analyze the methods for stability, and we highlight several examples of practical algorithms that give correct configurational statistics. Section~\ref{sec:velocities} associates possible velocity definitions with the derived St{\o}rmer-Verlet trajectories, and, based on specific statistical definitions of {\it on-site} and {\it half-step} velocities, we identify the statistically correct velocity expressions that properly match their spatial trajectories such that complete statistical information can be extracted from a simulation. Section~\ref{sec:combo} summarizes the complete set of developed algorithms into a single set of expressions that effectively captures all possibilities for conducting stochastic Verlet-type simulations of Langevin systems with both on-site and half-step velocity representations. Section~\ref{sec:other_velocities} briefly comments on features of other velocity expressions that represent neither on-site nor half-step relationships with the discrete-time spatial trajectory. Results of numerical validation of exemplified methods are shown in Section~\ref{sec:numerical_simulations} in order to demonstrate the applicability and statistical features of the methods as applied to both simple, one-dimensional nonlinear systems and complex Molecular Dynamics. Finally, in Section~\ref{sec:discussion}, we summarize the work.

%\tableofcontents

%%%%%%%%%%%%%%%%%%%%%%%%%%%%%%%%%%%%%%%%%%%%%%%%%%%%
\section{A brief review of the GJF and GJF-2GJ methods}
\label{sec:review}
In preparation for exploring the possible expressions for statistically correct methods, we here review the known statistically correct expressions for the GJF \cite{GJF1} and GJF-2GJ \cite{2GJ} methods, which will serve as the basis for our generalizations in the next sections. The discrete-time, finite-difference GJF equations that address the Langevin equation (\ref{eq:Langevin}) in the St{\o}rmer-Verlet (SV) form \cite{GJF2} are given as follows:
\begin{eqnarray}
r^{n+1} & = & 2br^n -ar^{n-1}+\frac{b\,dt^2}{m}f^n+\frac{b\,dt}{2m}(\beta^n+\beta^{n+1}) \; , \label{eq:gjf_sv_r}
\end{eqnarray}
where
\begin{subequations}
\begin{eqnarray}
a & = & \frac{\displaystyle{1-\frac{\alpha dt}{2m}}}{\displaystyle{1+\frac{\alpha dt}{2m}}} \label{eq:a} \\
b & = & \frac{\displaystyle{1}}{\displaystyle{1+\frac{\alpha dt}{2m}}}  \label{eq:b}
\end{eqnarray}
\end{subequations}
contain the linear friction coefficient $\alpha$.
The accumulated discrete-time noise during the time interval $(t_n,t_{n+1}]$ is
\begin{eqnarray}
\beta^{n+1} & = & \int_{t_n}^{t_{n+1}}\beta(t^\prime)\,dt^\prime \, , \label{eq:discrete_beta}
\end{eqnarray}
which results in an uncorrelated Gaussian random number with zero mean and a variance given by the temperature and friction coefficient:
\begin{subequations}
\begin{eqnarray}
\langle\beta^n\rangle & = & 0 \label{eq:noise_dis_ave} \\
\langle\beta^n\beta^l\rangle & = & 2\alpha k_BT \, dt \, \delta_{n,l} \, , \label{eq:noise_dis_std}
\end{eqnarray}
\label{eq:noise_dt}
\end{subequations}
where $\delta_{n,l}$ is Kronecker's delta function. The associated {\it on-site} GJF velocity variable $v^n$ at time $t_n$ is given \cite{2GJ} by
\begin{eqnarray}
v^n & = & \frac{r^{n+1}-(1-a)r^n-ar^{n-1}}{2b\,dt}+\frac{1}{4m}(\beta^n-\beta^{n+1}) \; , \label{eq:gjf_sv_v}
\end{eqnarray}
and the newly identified 2GJ {\it half-step} velocity $u^{n+\frac{1}{2}}$ at time $t_{n+\frac{1}{2}}$ is \cite{2GJ}
\begin{eqnarray}
u^{n+\frac{1}{2}} & = & \frac{r^{n+1}-r^n}{\sqrt{b}\,dt} \; , \label{eq:2gj}
\end{eqnarray}
where $\sqrt{b}\,u^{n+\frac{1}{2}} = v^{n+\frac{1}{2}}$ (Eq.~(\ref{eq:sv_u_basic})). Thus, the 2GJ half-step velocity $u^{n+\frac{1}{2}}$ is different from the standard half-step velocity $v^{n+\frac{1}{2}}$, the latter yielding statistically incorrect results for, e.g., the kinetic energy (see Ref.~\cite{2GJ}).

These equations can be written in three additional forms that all have identical outcomes for the trajectory $r^n$:\\
The velocity-Verlet (VV) GJF form \cite{GJF1} is
\begin{subequations}
\begin{eqnarray}
r^{n+1} & = & r^n+b\,\left[dt\,v^n+\frac{dt^2}{2m}f^n+\frac{dt}{2m}\beta^{n+1}\right] \label{eq:gjf_vv_r}\\
v^{n+1} & = & a\,v^n+\frac{dt}{2m}(af^n+f^{n+1})+\frac{b}{m}\beta^{n+1} \; ; \label{eq:gjf_vv_v}
\end{eqnarray}
\label{eq:gjf_vv}
\end{subequations}
the leap-frog (LF) GJF-2GJ form \cite{2GJ} is
\begin{subequations}
\begin{eqnarray}
u^{n+\frac{1}{2}} & = & a\,u^{n-\frac{1}{2}}+\frac{\sqrt{b}\,dt}{m}f^n+\frac{\sqrt{b}}{2m}(\beta^n+\beta^{n+1})\label{eq:LGJ2_v}\\
r^{n+1} & = & r^n+\sqrt{b}\,dt\,u^{n+\frac{1}{2}} \; ;  \label{eq:LGJ2_r}
\end{eqnarray}
\label{eq:gjf_lf}
\end{subequations}
and the compact GJF-2GJ form \cite{2GJ2} is
\begin{subequations}
\begin{eqnarray}
u^{n+\frac{1}{2}} & = & \sqrt{b}\;\left[v^n+\frac{dt}{2m}f^n+\frac{1}{2m}\beta^{n+1}\right] \label{eq:2GJ_u}\\
r^{n+1} & = & r^n+\sqrt{b}\,dt\,u^{n+\frac{1}{2}}\label{eq:2GJ_r}\\
v^{n+1} & = & \frac{a}{\sqrt{b}}u^{n+\frac{1}{2}}+\frac{dt}{2m}f^{n+1}+\frac{1}{2m}\beta^{n+1}\label{eq:2GJ_v} \, ,
\end{eqnarray}
\end{subequations}
which combines the VV and LF forms into one. In the remainder of this paper, we will refer to GJF, GJF-2GJ, and 2GJ as outlined in this section.

Linear analysis \cite{GJF1,2GJ} gives the following statistical results:\\
For the flat potential, $f=0$, the method yields the Einstein diffusion constant $D_E$ and the velocity autocorrelations:
\begin{eqnarray}
D_E & = & \lim_{n\,dt\rightarrow\infty}\frac{\langle (r^{q+n}-r^q)^2\rangle_q}{2\,n\,dt} \; = \; \frac{k_BT}{\alpha} \label{eq:gjf_stat_r}\\
\langle v^nv^{n}\rangle & = & \frac{k_BT}{m} \\
\langle u^{n+\frac{1}{2}}u^{n+\frac{1}{2}}\rangle & = & \frac{k_BT}{m} \; .
\end{eqnarray}
For the harmonic oscillator $f=-\kappa r$, with $\kappa>0$, we have:
\begin{eqnarray}
\langle r^nr^n\rangle & = & \frac{k_BT}{\kappa} \label{eq:gjf_stat_rr}\\
\langle v^nv^n\rangle & = & \frac{k_BT}{m}\left(1-\frac{\Omega_0^2dt^2}{4}\right) \label{eq:gjf_stat_vv}\\
\langle u^{n+\frac{1}{2}}u^{n+\frac{1}{2}}\rangle & = & \frac{k_BT}{m} \; , \label{eq:gjf_stat_uu}
\end{eqnarray}
where $\Omega_0=\sqrt{\kappa/m}$ is the natural frequency of the oscillator.
Equations (\ref{eq:gjf_stat_vv}) and (\ref{eq:gjf_stat_uu}) highlight the inherent time step dependence of the on-site velocity $v^n$ and the statistically robust, time-step-independent property of the half-step velocity $u^{n+\frac{1}{2}}$, which is the attractive feature of the GJF-2GJ method. Equations (\ref{eq:gjf_stat_r}) and (\ref{eq:gjf_stat_rr}) similarly show the robust time-step-independent configurational statistics, which is the attractive feature of the original GJF method. We note that the GJF configurational trajectory $r^n$ from Eq.~(\ref{eq:gjf_sv_r}) has been implemented in the Molecular Dynamics suite LAMMPS \cite{Plimpton,LAMMPS-Manual} as an option for a stochastic thermostat.

%%%%%%%%%%%%%%%%%%%%%%%%%%%%%%%%%%%%%%%%%%%%%%%%%%%%
\section{Developing the GJ stochastic St{\o}rmer-Verlet trajectories $r^n$}
\label{sec:sv_section}

Inspired by the SV expression for the spatial trajectory, Eq.~(\ref{eq:gjf_sv_r}), which shows that $r^n$ can be evaluated without any velocity variable for as long as contributions from the noise are included for the entire time span of the expression, namely $(t_{n-1},t_{n+1}]$, we explore possible statistically correct trajectories $r^n$ without concern for associated velocities. Associated velocities will be included in Sec.~\ref{sec:velocities}.

The general stochastic St{\o}rmer-Verlet form can be written
\begin{eqnarray}
r^{n+1} & = & 2c_1r^n-c_2r^{n-1}+c_3\frac{dt^2}{m}f^n+\frac{dt}{m}(c_4\beta^n+c_5\beta^{n+1}) \; , \label{eq:general_sv_r}
\end{eqnarray}
where $c_i$ are unit-less coefficients, which may depend on $\alpha$, $dt$, and $m$.  It is reasonable to require that $c_i$ must become the St{\o}rmer-Verlet coefficients in the frictionless limit shown in Eq.~(\ref{eq:sv_r_basic}); i.e., $c_1, c_2,c_3\rightarrow1$, and $c_4\beta^n, c_5\beta^{n+1}\rightarrow0$ for $\alpha\rightarrow0$. For $\alpha>0$ the coefficients will be determined by the statistical requirements on the trajectory $r^n$ for the harmonic oscillator $f=-\kappa r=-m\Omega_0^2r$ (for $\kappa>0$), which linearizes the above equation to read
\begin{eqnarray}
r^{n+1} & = & 2c_1Xr^n-c_2r^{n-1}+\frac{dt}{m}(c_4\beta^n+c_5\beta^{n+1}) \label{eq:general_sv_r_harm}\\
X & = & 1-\frac{c_3}{c_1}\frac{\Omega_0^2dt^2}{2} \; . \label{eq:X_1}
\end{eqnarray}
Multiplying Eq.~(\ref{eq:general_sv_r_harm}) with $r^{n-1}$, $r^n$, and $r^{n+1}$, and using, e.g., that $\langle r^nr^n\rangle=\langle r^{n+1}r^{n+1}\rangle$, we arrive at three coupled linear expressions of the correlations $\langle r^{n-1}r^{n+1}\rangle$,  $\langle r^{n}r^{n+1}\rangle$, $\langle r^{n}r^{n}\rangle$, expressed in the form
\begin{eqnarray}
\left(\begin{array}{ccc}
1 & -2c_1X & c_2 \\
0 & 1+c_2 & -2c_1X \\
c_2 & -2c_1X & 1 \end{array}\right)
\left(\begin{array}{c}
\langle r^{n-1} r^{n+1} \rangle \\
\langle r^{n}r^{n+1} \rangle \\
\langle r^{n}r^{n}\rangle\end{array}\right) & = &
\left(\begin{array}{c}
0 \\
c_4c_5 \\
c_4(2c_5c_1X+c_4)+c_5^2\end{array}\right)\frac{dt^2}{m^2}\langle\beta^n\beta^n\rangle\; ,  %\nonumber \\
\end{eqnarray}
which can be rewritten
\begin{eqnarray}
\left(\begin{array}{ccc}
1 & -2c_1X & c_2 \\
0 & 1+c_2 & -2c_1X \\
0 & -2c_1X & 1+c_2 \end{array}\right)
\left(\begin{array}{c}
\langle r^{n-1} r^{n+1} \rangle \\
\langle r^{n}r^{n+1} \rangle \\
\langle r^{n}r^{n}\rangle\end{array}\right) & = &
\left(\begin{array}{c}
0 \\
c_4c_5 \\
\frac{c_4(2c_5c_1X+c_4)+c_5^2}{1-c_2}\end{array}\right)8\frac{\alpha dt}{2m}\frac{c_1}{c_3}(1-X)\frac{k_BT}{\kappa} \; . \label{eq:system_correlations} %\nonumber \\
\end{eqnarray}
With some cumbersome, yet trivial calculations, we can find $\langle r^nr^n\rangle$ explicitly, and then require that the following expression is true for all relevant values of $\Omega_0dt$
\begin{eqnarray}
\langle r^nr^n\rangle & = & \underbrace{\left(\frac{8\frac{\alpha dt}{2m}\frac{c_4c_5}{c_3}}{1-c_2}\right)}_{=1}\overbrace{\left[\frac{(1-X)\left[\overbrace{\left(\frac{(1+c_2)(c_4^2+c_5^2)}{4c_4c_5c_1}\right)}^{=1}+X\right]}{\underbrace{\left(\frac{1+c_2}{2c_1}\right)^2}_{=1}-X^2}\right]}^{{\rm const}(=1)}\,\frac{k_BT}{\kappa} \; = \; \frac{k_BT}{\kappa} \; . %\nonumber \\
 \label{eq:cor_rr} 
\end{eqnarray}
For Eq.~(\ref{eq:general_sv_r}) to be a functional algorithm, the coefficients $c_i$ can depend on the known quantity $\alpha dt/2m$, but not on $\Omega_0dt$, which is not generally known for systems other than harmonic oscillators. As indicated in Eq.~(\ref{eq:cor_rr}), we must therefore require each of the two factors (fractions) to $k_BT/\kappa$ be unity (only $X$ depends on $(\Omega_0dt)^2$). Thus, the constraint given in Eq.~(\ref{eq:cor_rr}) can only be fulfilled if the following three conditions are fulfilled 
\begin{eqnarray}
\frac{2c_1}{1+c_2} & = & \pm1 \label{eq:constraint_c1c2}\\
(c_4\mp c_5)^2 & = & 0\label{eq:constraint_c4c5}\\
8\frac{\alpha dt}{2m}\frac{c_4c_5}{c_3} & = & 1-c_2 \; .\label{eq:constraint_c4c5c2c3}
\end{eqnarray}

In order to assess the ambiguity of the sign in Eq.~(\ref{eq:constraint_c1c2}), we write Eq.~(\ref{eq:general_sv_r}) in a velocity form, where $w_h^{n+\frac{1}{2}}$ is the traditional half-step velocity Eq.~(\ref{eq:sv_u_basic}) at time $t_{n+\frac{1}{2}}$
\begin{eqnarray}
\overbrace{\left(\frac{r^{n+1}-r^n}{dt}\right)}^{w_h^{n+\frac{1}{2}}} & = & c_2 \overbrace{\left(\frac{\frac{2c_1-1}{c_2}r^n-r^{n-1}}{dt}\right)}^{w_h^{n-\frac{1}{2}}}+\frac{c_3\,dt}{m}f^n+\frac{1}{m}(c_4\beta^n+c_5\beta^{n+1}) \; .\label{eq:sv_w}
\end{eqnarray}
We here see that any meaningful parameterization must require that the discrete-time velocity attenuation $c_2$ is limited by $|c_2|\le1$, and that only the positive sign in Eq.~(\ref{eq:constraint_c1c2}) is useful. Thus, ensuring that Eq.~(\ref{eq:cor_rr}) is satisfied, we have the three definite constraints
\begin{eqnarray}
2c_1 & = & 1+c_2 \label{eq:constraint_c1c2_final}\\
c_4 & = & c_5\label{eq:constraint_c4c5_final}\\
8\frac{\alpha dt}{2m}\frac{c_5^2}{c_3} & = & 1-c_2 \; , \label{eq:constraint_c2c3c5_final}
\end{eqnarray}
with $|c_2|<1$, implying that $c_1>0$ and $c_1\ge c_2$.

In addition to the spatial autocorrelation, Eq.~(\ref{eq:cor_rr}), we wish to ensure that diffusion for $f^n=0$ is in agreement with the continuous-time expectation $D_E=\frac{k_BT}{\alpha}$. Using the Einstein expression for diffusion, Eq.~(\ref{eq:gjf_stat_r}), we need to evaluate
\begin{eqnarray}
r^n-r^0 & = & dt\sum_{k=0}^{n-1}w_h^{k+\frac{1}{2}} \; , \label{eq:rn-r0}
\end{eqnarray}
where, inserting Eqs.~(\ref{eq:constraint_c1c2_final}) and (\ref{eq:constraint_c4c5_final}) into Eq.~(\ref{eq:sv_w}) for $f^n=0$, we write the velocity $w_h^{n+\frac{1}{2}}$ as a function of an initial value $w_h^{\frac{1}{2}}$ at time $t_{\frac{1}{2}}$
\begin{eqnarray}
w_h^{n+\frac{1}{2}} & = & c_2^nw_h^{\frac{1}{2}}+\frac{c_5}{m}\sum_{k=0}^{n-1}c_2^k(\beta^{n-k}+\beta^{n+1-k}) \; . \label{eq:wnw12}
\end{eqnarray}
Inserting Eq.~(\ref{eq:wnw12}) into Eq.~(\ref{eq:rn-r0}), retaining only the leading term in $n$ for $n\rightarrow\infty$, we obtain
\begin{eqnarray}
r^n-r^0 & \rightarrow & \chi \;  \frac{c_5}{m}\frac{dt}{1-c_2}\,2\,\sqrt{2\alpha k_BTdt\,n} \; , 
\end{eqnarray}
where $\chi\in N(0,1)$ is a stochastic number drawn from a normal distribution. The diffusion constant Eq.~(\ref{eq:gjf_stat_r}) is then evaluated to be
\begin{eqnarray}
D_E & = & \lim_{ndt\rightarrow\infty}\frac{(r^n-r^0)^2}{2\,dt\,n} \; = \; \left(\frac{\alpha dt}{2m}\frac{4c_5}{1-c_2}\right)^2\,\frac{k_BT}{\alpha} \, . 
\end{eqnarray}
It follows that 
\begin{eqnarray}
D_E & = &  \frac{k_BT}{\alpha} \; \; \Rightarrow \label{eq:D_E2} \\
\frac{1-c_2}{4c_5} & = & \pm\frac{\alpha dt}{2m} \; , \label{eq:constraint_c2c5_final}
\end{eqnarray}
where we, without loss of generality, can choose the positive sign on the right hand side of this expression. Combining the constraint Eq.~(\ref{eq:constraint_c2c5_final}) with Eq.~(\ref{eq:constraint_c2c3c5_final}) yields the simple relationship
\begin{eqnarray}
c_3 & = & 2c_5 \; . \label{eq:constraint_c3c5_final}
\end{eqnarray}

With the four developed constraints [Eqs.~(\ref{eq:constraint_c1c2_final}), (\ref{eq:constraint_c4c5_final}), (\ref{eq:constraint_c2c5_final}), and (\ref{eq:constraint_c3c5_final})] that arise from enforcing basic thermal statistics for a harmonic oscillator and diffusion on a flat surface [Eqs.~(\ref{eq:cor_rr}) and (\ref{eq:D_E2})], we write the resulting stochastic St{\o}rmer-Verlet trajectory Eq.~(\ref{eq:general_sv_r}) as a function of the single free functional parameter $c_2$:
\begin{eqnarray}
r^{n+1} & = & \overbrace{(1+c_2)}^{2c_1}r^n-c_2r^{n-1}+\overbrace{\left(\frac{1-c_2}{{\alpha dt/m}}\right)}^{c_3}\left[\frac{dt^2}{m}f^n+\frac{dt}{2m}(\beta^n+\beta^{n+1})\right] \; . \label{eq:GJ_sv_r}
\end{eqnarray}
{\bf We will refer to Eq.~(\ref{eq:GJ_sv_r}) as the GJ (Gr{\o}nbech-Jensen) set of SV trajectories, given by the parameter $c_2$, and the statistically correct sequence $r^n$ approximates the solution to Eq.~(\ref{eq:Langevin})}. The functional parameter $c_2$ (a function of $\frac{\alpha dt}{m}$) is the one-time-step velocity attenuation factor [see Eq.~(\ref{eq:sv_w})], generally restricted by $|c_2|<1$. Since we must require that $c_3\rightarrow1$ for $\alpha dt/m\rightarrow0$, it is implied that $c_2$ is a monotonically decaying function with the limiting form $c_2\rightarrow1\!-\!\alpha dt/m$ for $\alpha dt/m\rightarrow0$. 

We notice that the terminal drift for constant external force $f$ is correctly given by
\begin{eqnarray}
\langle r^n-r^{n-1}\rangle & = & \frac{f}{\alpha}\,dt \label{eq:r_drift}
\end{eqnarray}
for any choice $|c_2|<1$.

Table~\ref{tab:sv} summarizes the relationships between the five parameters that determine the set of GJ methods in the SV form derived in this section. We notice that the two known configurationally correct methods \cite{GJF1,Sivak} are special cases of the GJ set of possibilities. We highlight these two cases, along with a few others, in Sec.~\ref{sec:sv_highlight} below.
\begin{table}
{{{\begin{tabular}{|ccccc||c|}\hline
& & Expressed by: & & & \\
\makebox[0.85 in]{$c_1$} &&  \makebox[0.85 in]{$c_2$} && \makebox[0.85 in]{$c_3$} & Properties \\ & & & & &  \\ \hline\hline
& & &  & & $ 0<c_1\le1$ \\
$c_1$ & = & $\displaystyle\frac{1+c_2}{2}$  & = & $\displaystyle\frac{1}{2}(\displaystyle\frac{\alpha dt}{2m}c_3-1)$ & $\displaystyle\lim_{\alpha dt\rightarrow0}c_1=1$ \\ & & & & & \\ \hline
& & & & &$|c_2|\le1$  \\
$2c_1-1$ & = & $c_2$ & = & $1-\displaystyle\frac{\alpha dt}{m}c_3$ & $c_2\rightarrow1-\displaystyle\frac{\alpha dt}{m}$  \\ & & & & & for $\alpha dt \rightarrow0$ \\ \hline
& & & & & $0<c_3$  \\
$\displaystyle\frac{2m}{\alpha dt}(1-c_1)$ & = & $\displaystyle\frac{m}{\alpha dt}(1-c_2)$ & = & $c_3\;=\;2c_4\;=\;2c_5$ & $\displaystyle\lim_{\alpha dt\rightarrow0}c_3=1$ \\   & & & & & \\ \hline \end{tabular}
}}}
\caption{Overview of the relationships between the five parameters defining the general GJ trajectory, Eq.~(\ref{eq:GJ_sv_r}), which has one free functional parameter. The velocity attenuation factor during one time step is $c_2$.}
\label{tab:sv}
\end{table}

%%%%%%%%%%%%%%%%%%%%%%%%%%%%%%%%%%%%%%%%%%%%%%%%%%%%
\subsection{Stability of the GJ trajectories}
\label{sec:stability}
For $k_BT=0$ we insert a Hooke's force $f^n=-\kappa r^n=-m\Omega_0^2r^n$ into Eq.~(\ref{eq:GJ_sv_r}) to obtain
\begin{eqnarray}
r^{n+1} & = & 2c_1Xr^n-c_2r^{n-1} \; , \label{eq:Hooke_GJ_sv_r}%\\
\end{eqnarray}
where $X$ is given by Eq.~(\ref{eq:X_1}). The roots $\Lambda_\pm$ of the characteristic polynomial are
\begin{eqnarray}
\Lambda_\pm & = & c_1X\pm\sqrt{c_1^2X^2-c_2}\; , \label{eq:eigenvalues}
\end{eqnarray}
and stability is given when $|\Lambda_\pm|\le1$. We separate the two cases of real and complex eigenvalues:\\

\noindent
{\underline{$\Lambda_\pm$ complex}}: When
\begin{eqnarray}
0 & \le & c_1^2X^2 \; < \; c_2\; , \label{eq:underdamped}
\end{eqnarray}
the dynamics is underdamped, and $|\Lambda_\pm|=\sqrt{c_2}<1$ implies that the trajectory is stable. For the frictionless case, $\alpha=0$ ($c_1=c_2=1$), we recover the usual Verlet stability range $\Omega_0dt<2$.

In this stable regime of complex eigenvalues $\Lambda_\pm$, the eigenvalues can be written
\begin{eqnarray}
\Lambda_\pm & = & \sqrt{c_2}\,e^{\pm i\Omega_Vdt} \; , 
\end{eqnarray}
where $\Omega_V$ is the discrete-time oscillator frequency found from Eq.~(\ref{eq:eigenvalues}) to be given by
\begin{subequations}
\begin{eqnarray}
\sqrt{c_2}\,\cos\Omega_Vdt & = & c_1\left(1-\frac{c_3}{c_1}\frac{\Omega_0^2dt^2}{2}\right) \\
\sqrt{c_2}\,\sin\Omega_Vdt & = & \Omega_0dt\,\sqrt{c_1c_3}\,\sqrt{1-\frac{c_3}{c_1}\left[\left(\frac{\Omega_0dt}{2}\right)^2+\left(\frac{\alpha}{2m\Omega_0}\right)^2\right]} \; .
\end{eqnarray}
\label{eq:WV}
\end{subequations}

\noindent
{\underline{$\Lambda_\pm$ real}}: When
\begin{eqnarray}
c_2 & < & c_1^2X^2 \; , 
\end{eqnarray}
stability requires that $-1<\Lambda_\pm<1$.\\
We first assume that $X>0$. In this case we have the two conditions
\begin{eqnarray}
-1 & < & \Lambda_- \; \; \Rightarrow \; \; 0 \; < \; 2c_1(1+X) \\
\Lambda_+ & < & 1 \; \; \Rightarrow \; \; X \; < \; 1 \; < \; \frac{1}{c_1} \; .
\end{eqnarray}
Thus, stability implies that $X<1$.\\
We then assume that $X<0$. In this case we have the two conditions
\begin{eqnarray}
-1 & < & \Lambda_- \; \; \Rightarrow \; \; |X| \; < \; 1 \; < \; \frac{1}{c_1} \\
\Lambda_+ & < & 1 \; \; \Rightarrow \; \; 0 \; < \; 2c_1(1+|X|) \; , 
\end{eqnarray}
which again shows that stability is implied for $|X|<1$.\\
In summary, the stability criterion is $|\Lambda_\pm|<1\Leftrightarrow|X|<1$; i.e., that
\begin{eqnarray}
\Omega_0^2dt^2 & < & 4\frac{\alpha dt/m}{1-c_2}\frac{1+c_2}{2} \; = \; 4\frac{c_1}{c_3} \; , \label{eq:Stability}
\end{eqnarray}
where, as mentioned above, $c_2$ is a function of $\frac{\alpha dt}{m}$ such that $c_2\rightarrow1\!-\!\alpha dt/m$ for $\alpha dt/m\rightarrow0$.
Notice that the Verlet stability criterion for $\alpha=0$, $\Omega_0dt\le2$, is always satisfied if $c_2\ge a\Leftrightarrow c_1\ge c_3$.

%%%%%%%%%%%%%%%%%%%%%%%%%%%%%%%%%%%%%%%%%%%%%%%%%%%%
\subsection{Examples of specific GJ trajectories}
\label{sec:sv_highlight}
\begin{figure}[t]
\centering
\scalebox{0.8}{\centering \includegraphics[trim={1.5cm 7.0cm 1cm 2.0cm},clip]{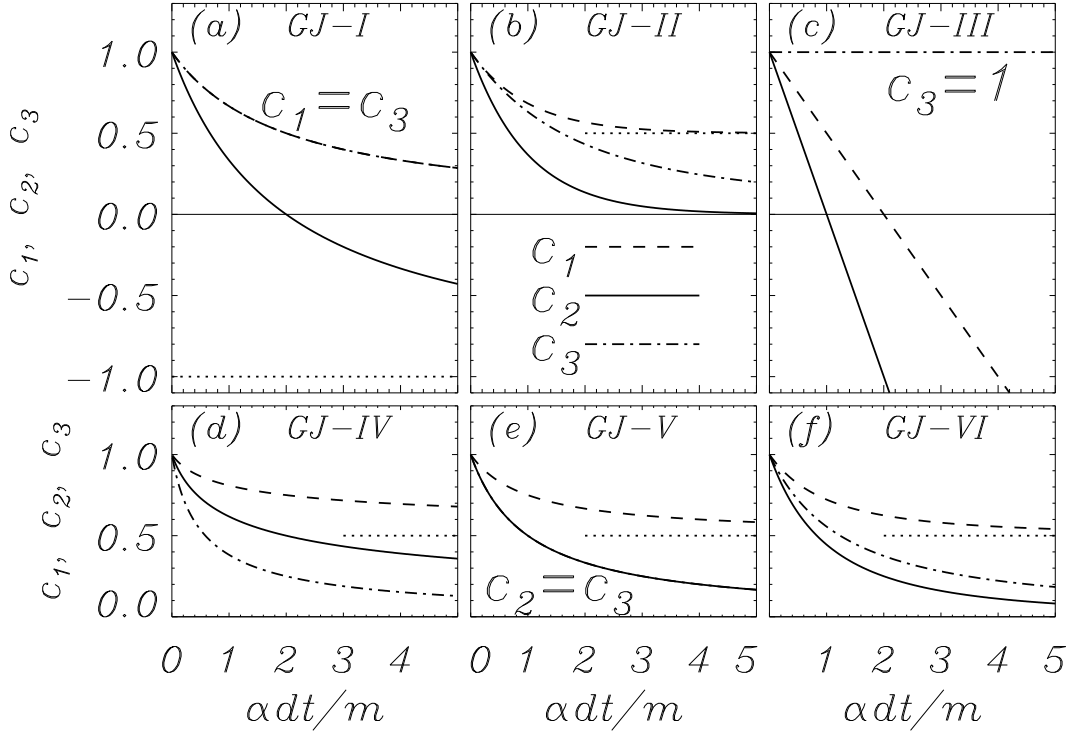}}
%\vspace{-0.50 cm}
\caption{GJ trajectory coefficients for the highlighted inertial cases I-VI in (a-f). Linestyles for each coefficient are given in (b). Horizontal dotted lines indicate: (a) asymptote $c_2\rightarrow-1$ for $\alpha dt/m\rightarrow\infty$; (b, d, e, f) asymptotic value $c_1\rightarrow\frac{1}{2}$ for $\alpha dt/m\rightarrow\infty$ in cases II, IV, V, and VI. In these cases $c_2, c_3\rightarrow0$ for $\alpha dt/m\rightarrow\infty$. In (a) $c_1=c_3\rightarrow0$, while $c_2\rightarrow-1$ for $\alpha dt/m\rightarrow\infty$.
}
\label{fig_1}
\end{figure}

\begin{figure}[t]
\centering
\scalebox{0.8}{\centering \includegraphics[trim={1.5cm 6.5cm 1.5cm 4.0cm},clip]{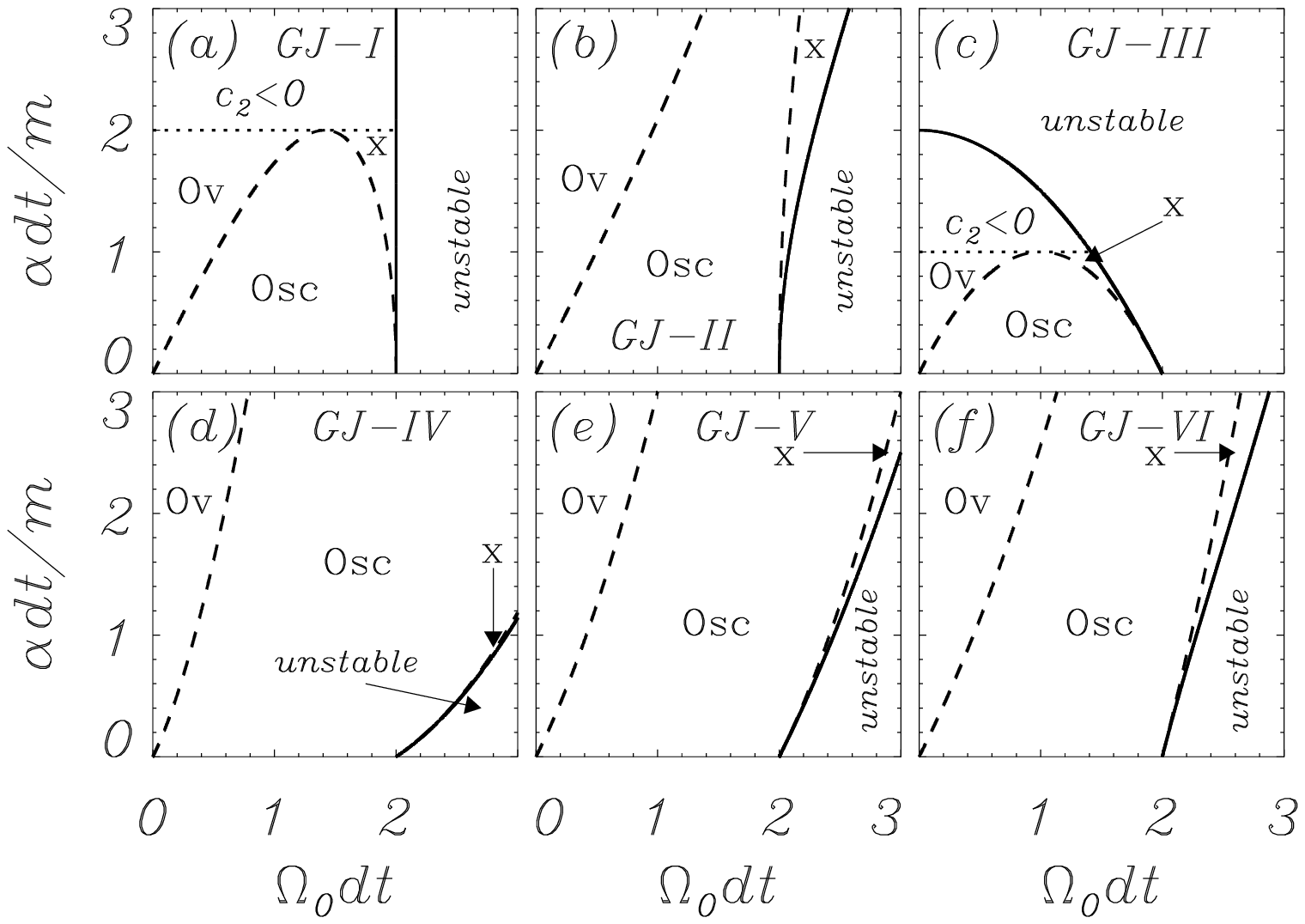}}
%\vspace{-0.50 cm}
\caption{Stability properties of the highlighted inertial GJ methods I-VI in (a-f) for which the coefficients are shown in Fig.~\ref{fig_1}. Solid curves indicate boundaries between stable and unstable behavior. Stable regions, given by Eq.~(\ref{eq:Stability}), are separated into underdamped oscillations (Osc, as given by Eq.~(\ref{eq:underdamped}), enclosed by dashed curves), overdamped dynamics (Ov, defined by $0<\Lambda_\pm<1$), and the region (x), which is where $-1<\Lambda_\pm<0$. This region is an unphysical, large time step region for which the frequency is $\Omega_V=\pi/dt$. Methods I and III, (a) and (c), also have a stable, unphysical region for $c_2<0$ $\Leftrightarrow$ $\Lambda_-\Lambda_+<0$ (see Eq.~(\ref{eq:eigenvalues})).
}
\label{fig_2}
\end{figure}
We here give several specific examples of GJ methods, which by design all sample correct Boltzmann statistics for a linear system regardless of the time step within the linear stability limit given by Eq.~(\ref{eq:Stability}).
Since the GJ set of methods, Eq.~(\ref{eq:GJ_sv_r}), given by the one-time-step velocity attenuation factor $c_2$, is the only possible form of statistically correct stochastic Verlet-type trajectories, we must validate that existing known methods with correct linear statistics are included in this set of methods.\\
\begin{description}
\item[{\underline{GJ-I}}:] The GJF trajectory \cite{GJF1,GJF2} shown in Eq.~(\ref{eq:gjf_sv_r}) is found for
\begin{eqnarray}
c_2 & = & a \; , 
\end{eqnarray}
where $a$ is given in Eq.~(\ref{eq:a}), and $2c_1=2c_3=1+a$. The linear stability limit is given by Eq.~(\ref{eq:Stability}):
\begin{eqnarray}
\Omega_0dt & < & 2\; .
\end{eqnarray}
The trajectory exhibits unphysical temporal oscillations for $\alpha dt > 2m$. See Fig.~\ref{fig_1}a for a visual representation of the coefficients $c_1$, $c_2$, and $c_3$, and Fig.~\ref{fig_2}a for a diagram of the stability properties.

\item[{\underline{GJ-II}}:] The trajectory shown in Ref.~\cite{Sivak}, built on Ref.~\cite{ML}, is found for
\begin{eqnarray}
c_2 & = & \exp(-\frac{\alpha dt}{m}) \; . \label{eq:c2_case-B}
\end{eqnarray}
Since $c_2\ge a$ for all $\alpha dt/m$, we conclude that the trajectory is linearly stable for at least all $\Omega_0dt<2$. See Fig.~\ref{fig_1}b for a visual representation of the coefficients $c_1$, $c_2$, and $c_3$, and Fig.~\ref{fig_2}b for a diagram of the stability properties.
\end{description}
Thus, the two known configurationally correct methods are contained in the GJ set of methods. The relationship between these two methods has also been pointed out in Ref.~\cite{Li}. Additionally, we exemplify other cases of trajectories in this set of algorithms that may have interest:
\begin{description}
\item[{\underline{GJ-III}}:] $c_2=1-\frac{\alpha dt}{m}$, which implies that $c_3=1$. This case yields the trajectory:
\begin{eqnarray}
r^{n+1} & = & (2-\frac{\alpha dt}{m})r^n-(1-\frac{\alpha dt}{m})r^{n-1}+\frac{dt^2}{m}f^n+\frac{dt}{2m}(\beta^n+\beta^{n+1}) \; .\label{eq:GJ_sv_C}
\end{eqnarray}
The method is characterized by not directly scaling the force and noise terms with a function of $\alpha dt/m$ in Eq.~(\ref{eq:GJ_sv_r}). 
The method is of particular interest for lightly damped systems, due to observations made in Sec.~\ref{sec:combo}. The linear stability range is given by Eq.~(\ref{eq:Stability}) to
\begin{eqnarray}
\Omega_0dt & < & \sqrt{4+\left(\frac{\alpha}{m\Omega_0}\right)^2}-\frac{\alpha}{m\Omega_0} \; . \label{eq:Stability_III}
\end{eqnarray}
See Fig.~\ref{fig_1}c for a visual representation of the coefficients $c_1$, $c_2$, and $c_3$, and Fig.~\ref{fig_2}c for a diagram of the stability properties.

\item[{\underline{GJ-IV}}:] $c_2=\sqrt{c_3}$. In this case we have
\begin{eqnarray}
c_2 & = & \sqrt{c_3} \; = \; \frac{\sqrt{1+4\frac{\alpha dt}{m}}-1}{2\frac{\alpha dt}{m}} \; .
\end{eqnarray}
Since $c_2>a$, the linear stability limit exceeds the usual Verlet criterion $\Omega_0dt<2$. This relationship between $c_2$ and $c_3$ may be attractive for reasons given in Sec.~\ref{sec:combo}. See Fig.~\ref{fig_1}d for a visual representation of the coefficients $c_1$, $c_2$, and $c_3$, and Fig.~\ref{fig_2}d for a diagram of the stability properties.

\item[{\underline{GJ-V}}:] $c_2=c_3$. In this case we have
\begin{eqnarray}
c_2 & = & c_3 \; = \; \frac{1}{1+\frac{\alpha dt}{m}} \\
c_1 & = & 2\frac{c_2}{b} \; . 
\end{eqnarray}
The resulting trajectory is given by:
\begin{eqnarray}
r^{n+1} & = & c_2\left[\frac{2}{b}r^n-r^{n-1}+\frac{dt^2}{m}f^n+\frac{dt}{2m}(\beta^n+\beta^{n+1})\right] \; . \label{eq:GJ_sv_E}
\end{eqnarray}
Since $c_2>a$, the linear stability limit exceeds the usual Verlet criterion $\Omega_0dt<2$. See Fig.~\ref{fig_1}e for a visual representation of the coefficients $c_1$, $c_2$, and $c_3$, and Fig.~\ref{fig_2}e for a diagram of the stability properties.

\item[{\underline{GJ-VI}}:] $c_2=b^2$, where $b$ is given in Eq.~(\ref{eq:b}). This case yields the trajectory:
\begin{eqnarray}
r^{n+1} & = & (1+b^2)r^n-b^2r^{n-1}+b\frac{1+b}{2}\left[\frac{dt^2}{m}f^n+\frac{dt}{2m}(\beta^n+\beta^{n+1})\right] \; . \label{eq:GJ_sv_D}
\end{eqnarray}
Since $c_2=b^2>a$ the linear stability limit exceeds the usual Verlet criterion $\Omega_0dt<2$. See Fig.~\ref{fig_1}f for a visual representation of the coefficients $c_1$, $c_2$, and $c_3$, and Fig.~\ref{fig_2}f for a diagram of the stability properties.

\item[{\underline{GJ-0}}:] It is apparent from Eq.~(\ref{eq:GJ_sv_r}) that the mass does not play an explicit role in the method. We may therefore, as also noted in Ref.~\cite{Sivak} for Eq.~(\ref{eq:c2_case-B}), choose $c_2=0$ and write the equation
\begin{eqnarray}
r^{n+1} & = & r^n+\frac{1}{\alpha}\left[dt\,f^n+\frac{1}{2}(\beta^n+\beta^{n+1})\right] \; , \label{eq:GJ_sv_r2}
\end{eqnarray}
which is an explicit, statistically correct, discrete-time Euler approximation to the solution of the massless Brownian dynamics equation
\begin{eqnarray}
\alpha\dot{r} & = f+\beta \; . \label{eq:Brownian}
\end{eqnarray}
The linear stability limit is given by Eq.~(\ref{eq:Stability}) to be the expected Euler limit
\begin{eqnarray}
\frac{\kappa}{\alpha}dt & < & 2 \; . \label{eq:stability_Brownian}
\end{eqnarray}
Notice that this method for Brownian behavior is reached from any of the above exemplified methods by choosing $\alpha dt/m=2$ (GJ-I), $\alpha dt/m\rightarrow\infty$ (GJ-II,IV,V,VI), and $\alpha dt/m=1$ (GJ-III).

\end{description}

Among the inertial methods, GJ-II, IV-VI are qualitatively equivalent in that $c_1\rightarrow\frac{1}{2}$ and $c_3\rightarrow0$ for $\alpha dt/m\rightarrow\infty$ and that $c_1/c_3>1$. The stability diagrams in Fig.~\ref{fig_2} further support this qualitative similarity leaving an impression that for $\alpha>0$ GJ-IV has a wider stability range than GJ-V, which is more stable than GJ-VI and II. However, this does not necessarily mean that the efficiency of sampling the phase space using GJ-IV is better at larger $dt$ than, e.g., GJ-II, since, e.g.,  the discrete-time harmonic frequency $\Omega_V$ in the underdamped regime (see Eq.~(\ref{eq:WV})) is not the same for all methods. In fact, the frequency $\Omega_V$ is given by
\begin{eqnarray}
\tan{\Omega_Vdt} & = & \sqrt{\frac{c_3}{c_1}}\Omega_0dt\frac{\sqrt{1-\left(\sqrt\frac{c_3}{c_1}\frac{\Omega_0dt}{2}\right)^2-\left(\sqrt{\frac{c_3}{c_1}}\frac{\alpha}{2m\Omega_0}\right)^2}}{1-\left(\sqrt\frac{c_3}{c_1}\frac{\Omega_0dt}{2}\right)^2} \; ,
\end{eqnarray}
which implies that the time evolution is largely given by the scaled frequency $\sqrt{c_3/c_1}\Omega_0$. Thus, the apparent differences in stability ranges, seen in Fig.~\ref{fig_2} and quantified in Eq.~(\ref{eq:Stability}), mostly reflect that the time evolution of the system is scaled differently for the different methods. We notice that GJ-I (the GJF method) is the unscaled case.

In the massless limit (GJ-0), both the stability range Eq.~(\ref{eq:stability_Brownian}) and the sampling of the phase-space are given solely by $dt/\alpha$, and an increase in $\alpha$ does therefore not lead to a real increase in sampling efficiency since the dynamics slows down and $dt$ needs to be scaled accordingly.
%%%%%%%%%%%%%%%%%%%%%%%%%%%%%%%%%%%%%%%%%%%%%%%%%%%%
\section{Accompanying velocities to the GJ trajectories}
\label{sec:velocities}

Having identified the complete set of methods that will give correct time-step-independent configurational statistics, we start similarly to Ref.~\cite{GJF-2} for identifying possible finite difference velocities to accompany the trajectories from Sec.~\ref{sec:sv_section}. In a similar manner to Eq.~(\ref{eq:gjf_sv_r}) leading to the general expression Eq.~(\ref{eq:general_sv_r}), we draw visual cues from Eq.~(\ref{eq:gjf_sv_v}) to explore the general finite difference expression for a velocity $w^\nu$ at time $t_\nu$
\begin{eqnarray}
  w^\nu & = & \frac{\gamma_1r^{n+1}+\gamma_2r^n+\gamma_3r^{n-1}}{dt}
  +\frac{\gamma_4\beta^n+\gamma_5\beta^{n+1}}{m} \; , 
  \label{eq:GJF-2GJF}
\end{eqnarray}
where $\gamma_i$ are unit-less coefficients (which may be functions of $\frac{\alpha dt}{m}$) to be determined by statistical requirements, and where $\nu\in[n-1,n+1]$ refers to some time, where the finite difference expression coincides with the actual velocity. We immediately make the observation that
\begin{eqnarray}
\gamma_1+\gamma_2+\gamma_3 & = & 0 \label{eq:gamma_cond_1}
\end{eqnarray}
in order to ensure that Eq.~(\ref{eq:GJF-2GJF}) is a finite difference approximation to a differential.
The two noise terms, $\beta^n$ and $\beta^{n+1}$, span the
time interval of the finite difference, $t_{n-1}<t<t_{n+1}$, which defines the velocity $w^\nu$ (see Eq.~(\ref{eq:discrete_beta})).
Notice that this undetermined velocity $w^\nu$ can describe the usual zero-friction, finite difference expressions somewhere in the interval $t_{n-1}\le t \le t_{n+1}$, including the on-site velocity Eq.~(\ref{eq:sv_v_basic}) at $t_n$, as
\begin{eqnarray}
w^n & = & \frac{r^{n+1}-r^{n-1}}{2dt} \; , \nonumber
\end{eqnarray}
for $\gamma_1=-\gamma_3=\frac{1}{2}$ and $\gamma_2=\gamma_4=\gamma_5=0$, and the half-step velocity Eq.~(\ref{eq:sv_u_basic}) at $t_{n+\frac{1}{2}}$
\begin{eqnarray}
w^{n+\frac{1}{2}} & = & \frac{r^{n+1}-r^n}{dt} \; , \nonumber
\end{eqnarray}
for $\gamma_1=-\gamma_2=1$ and $\gamma_3=\gamma_4=\gamma_5=0$.

For the general velocity $w^\nu$ at a certain time $t_\nu$, we can immediately identify a general parameter constraint by
inserting Eq.~(\ref{eq:GJ_sv_r}) into Eq.~(\ref{eq:GJF-2GJF}) to obtain the velocity $w^{\nu-1}$ such that the expression displays the velocity attenuation correctly over a time step:
\begin{eqnarray}
w^\nu & = & \frac{(\gamma_1c_2-\gamma_3)r^n+(\gamma_3-\gamma_1c_2)r^{n-1}}{dt}+\frac{(\gamma_4+c_4\gamma_1)\beta^n+(\gamma_5+c_5\gamma_1)\beta^{n+1}}{m} %\nonumber \\
%&&
+c_3\gamma_1\frac{dt}{m}f^n \nonumber \\ \\
& = & c_2\Big[\overbrace{\frac{\gamma_1r^n+\gamma_2r^{n-1}+\gamma_3r^{n-2}}{dt}-\underbrace{\left(\frac{c_4\gamma_3}{c_2m}\right)}_{-\gamma_4/m}\beta^{n-1}+\frac{\gamma_5}{m}\beta^{n}}^{w^{\nu-1}}\Big]\nonumber \\
&&+\frac{\gamma_1c_4-\gamma_3c_5+\gamma_4-c_2\gamma_5}{m} \beta^n+\frac{\gamma_1c_5+\gamma_5}{m}\beta^{n+1}+c_3\frac{dt}{m}(\gamma_1f^{n}-\gamma_3f^{n-1}) \\
& = & c_2w^{\nu-1}+\frac{(\gamma_1-\gamma_3)c_5+\gamma_4-c_2\gamma_5}{m} \beta^n+\frac{\gamma_1c_5+\gamma_5}{m}\beta^{n+1}+c_3\frac{dt}{m}(\gamma_1f^{n}-\gamma_3f^{n-1}) \; . % \nonumber \\
  \label{eq:recursion_1} 
\end{eqnarray}
Thus, we conclude that
\begin{eqnarray}
c_2\gamma_4 & = & -c_4\gamma_3 \; = \; -\frac{1}{2}c_3\gamma_3\, , \label{eq:gamma_cond_2}
\end{eqnarray}
since this allows for an expression that applies to all time steps, with $c_2$ being the velocity attenuation factor identified in Eq.~(\ref{eq:sv_w}).

Using the general expression for the velocity $w^\nu$ we write the velocity autocorrelation, which must be both time-step-independent and correct for a velocity at equilibrium:
\begin{eqnarray}
\langle w^\nu w^\nu\rangle & = &
\frac{\gamma_1^2+\gamma_2^2+\gamma_3^2}{dt^2}\langle
(r^n)^2\rangle+2\frac{\gamma_1\gamma_2+\gamma_2\gamma_3}{dt^2}
\langle r^nr^{n+1}\rangle \nonumber \\
& + & 2\frac{\gamma_1\gamma_3}{dt^2}\langle r^{n-1}r^{n+1}\rangle +
\frac{\gamma_4^2+\gamma_5^2}{m^2}\langle\beta^n\beta^n\rangle\nonumber\\
& + & 2\frac{\gamma_1\gamma_4}{m\,dt}\langle
r^{n+1}\beta^n\rangle+2\frac{\gamma_1\gamma_5+\gamma_2\gamma_4}{m\,dt}\langle
r^n\beta^n\rangle\; = \; \frac{k_BT}{m} \, , \label{eq:w2_1}
\end{eqnarray}
where, for a harmonic oscillator, we see from Eq.~(\ref{eq:GJ_sv_r}) that
\begin{eqnarray}
\langle r^{n+1}\beta^n\rangle & = & 2c_1X\langle r^n\beta^n\rangle+\frac{dt}{2m}c_3\langle\beta^n\beta^n\rangle \label{eq:rnp1betan}\\
\langle r^n\beta^n\rangle & = & \frac{dt}{2m}c_3\langle\beta^n\beta^n\rangle \; , \label{eq:rnbetan}
\end{eqnarray}
and where $\langle\beta^n\beta^n\rangle$ is given by Eq.~(\ref{eq:noise_dis_std}).

Using the results from Sec.~\ref{sec:sv_section}, where the relationships between the coefficients $c_i$ are summarized in Table~\ref{tab:sv}, Eq.~(\ref{eq:system_correlations}) simplifies considerably
\begin{eqnarray}
&& \left(\begin{array}{ccc}
1 & -2c_1X & c_2 \\
0 & 1 & -X \\
0 & 0 & 1 \end{array}\right)
\left(\begin{array}{c}
\langle r^{n-1} r^{n+1} \rangle \\
\langle r^{n}r^{n+1} \rangle \\
\langle r^{n}r^{n}\rangle\end{array}\right) \; = \;
\left(\begin{array}{c}
0 \\
\frac{1}{2}(1-c_2)(1-X) \\
1\end{array}\right)\frac{k_BT}{\kappa} %\nonumber  \\
 \\
&&
\Rightarrow \; \; \left(\begin{array}{c}
\langle r^{n-1} r^{n+1} \rangle \\
\langle r^{n}r^{n+1} \rangle \\
\langle r^{n}r^{n}\rangle\end{array}\right) \; = \;
\left(\begin{array}{c}
-c_2 +\frac{1}{2}X\left[(1-c_2^2)+(1+c_2)^2X\right]\\
\frac{1}{2}\left[(1-c_2)+(1+c_2)X\right] \\
1\end{array}\right)\frac{k_BT}{\kappa} \label{eq:auto-correlations} \; . %\nonumber \\
\end{eqnarray}
Inserting the correlations, Eqs.~(\ref{eq:rnp1betan}),  (\ref{eq:rnbetan}), and (\ref{eq:auto-correlations}), into the velocity auto correlation function, Eq.~(\ref{eq:w2_1}), yields the expression
\begin{eqnarray}
\langle w^\nu w^\nu \rangle & = &\left[ (\Omega_0^2dt^2)^{-1}\,\Gamma_{-1}+\Gamma_0+(\Omega_0^2dt^2)\,\Gamma_1\right] \, \frac{k_BT}{m} \; = \; \frac{k_BT}{m} \; , \label{eq:w2_2}
\end{eqnarray}
where the three coefficients, which may depend on $\alpha dt/m$ (but not on $(\Omega_0dt)$), are given by:
\begin{subequations}
\begin{eqnarray}
\Gamma_{-1}(\frac{\alpha dt}{m}) & = & \gamma_1^2+\gamma_2^2+\gamma_3^2+2\gamma_1\gamma_2+2\gamma_2\gamma_3+2\gamma_1\gamma_3 \; = \; (\gamma_1+\gamma_2+\gamma_3)^2\label{eq:Gamma_m1}\\
\Gamma_0 (\frac{\alpha dt}{m})& = & c_3\left[2\frac{\alpha dt}{m}(2c_1+1)\gamma_1\gamma_4-\gamma_2(\gamma_1+\gamma_3)-2(1+c_1)\gamma_1\gamma_3\right] \nonumber \\
&& + \; 2\frac{\alpha dt}{m}(\gamma_4^2+\gamma_5^2+c_3(\gamma_1\gamma_5+\gamma_2\gamma_4)) \label{eq:Gamma_0}\\
\Gamma_1(\frac{\alpha dt}{m}) & = & c_3^2\gamma_1(\gamma_3-2\frac{\alpha dt}{m}\gamma_4) \; . \label{eq:Gamma_1}
\end{eqnarray}
\label{eq:Gamma_all}
\end{subequations}
Thus, Eq.~(\ref{eq:w2_2}) displays three constraints for determining $\gamma_i$ such that the velocity exhibits the correct equipartition for all time steps. The first constraint, $\Gamma_{-1}=0$, is redundant with Eq.~(\ref{eq:gamma_cond_1}), and the two others are
\begin{subequations}
\begin{eqnarray}
\Gamma_{0}(\frac{\alpha dt}{m}) & = & 1 \label{eq:condition_G0}\\
\Gamma_{1}(\frac{\alpha dt}{m}) & = & 0 \; . \label{eq:condition_G1}
\end{eqnarray}
\end{subequations}
Notice that Eq.~(\ref{eq:condition_G1}) is generally in conflict with Eq.~(\ref{eq:gamma_cond_2}) unless $\gamma_3=\gamma_4=0$.

Since the determination of $\gamma_i$ is not immediately transparent from these expressions, we will comment on some known results for the GJF method ($c_2=a$) before proceeding.

The GJF on-site velocity, Eq.~(\ref{eq:gjf_sv_v}), can be expressed by the values: $\gamma_1=1/2b$, $\gamma_2=-(1-a)/2b$, $\gamma_3=-(\gamma_1+\gamma_2)$, and $\gamma_4=-\gamma_5=\frac{1}{4}$. Inserting these coefficients into Eq.~(\ref{eq:Gamma_all}) gives $\Gamma_{-1}=0$, $\Gamma_0=1$, and $\Gamma_1=-\frac{1}{2}$, in agreement with the known result, Eq.~(\ref{eq:gjf_stat_vv}). Thus, as pointed out in Ref.~\cite{GJF-2} for the GJF method (GJ-I), the on-site velocity is troubled by the relatively simple relationship given by Eqs.~(\ref{eq:Gamma_1}) and (\ref{eq:condition_G1}).

The 2GJ half-step velocity \cite{2GJ} (given in Eq.~(\ref{eq:2gj})) can be expressed by the values: $\gamma_1=-\gamma_2=1/\sqrt{b}$,  $\gamma_3=\gamma_4=\gamma_5=0$. Inserting these coefficients into Eqs.~(\ref{eq:Gamma_all}) gives $\Gamma_{-1}=0$, $\Gamma_0=1$, and $\Gamma_1=0$, which is precisely the desired outcome. Thus, as pointed out in Ref.~\cite{2GJ}, the half-step velocity is statistically correct regardless of the time step.

In light of the velocities relevant for the GJF method, we will now provide two types of suggestions for statistical definitions of what is meant by {\it on-site} and {\it half-step} velocities. First, in Sec.~\ref{sec:on-site}, we give a general definition of what it means to be an on-site velocity, and determine the resulting properties of possible on-site velocities that can accompany GJ trajectories. Second, in Sec.~\ref{sec:half-step}, we give a general definition of what it means to be a half-step velocity, and generalize the result of Ref.~\cite{2GJ} to produce statistically correct half-step velocities that can accompany the set of GJ trajectories.

%%%%%%%%%%%%%%%%%%%%%%%%%%%%%%%%%%%%%%%%%%%%%%%%%%%%
\subsection{On-site velocity $v^{n}$}
\label{sec:on-site}
An on-site velocity is well defined for $\alpha=0$, since it can be symmetrically expressed by the central difference given in Eq.~(\ref{eq:sv_v_basic}). However, when $\alpha dt>0$, it may become less obvious from the finite difference expression which precise time the expression should be associated with. For example, the GJF on-site velocity $v^n$ given in Eq.~(\ref{eq:gjf_sv_v}) is symmetry broken by the friction, and for $\alpha dt = 2m$ ($a=0$) this expression appears to be a symmetric finite difference centered around the time $t_{n+\frac{1}{2}}$. Thus, the finite difference expression itself only indicates that it represents a velocity somewhere in the interval $t\in[t_{n-1},t_{n+1}]$. We therefore need to make a reasonable definition that can clarify the identity of an on-site velocity for $\alpha dt>0$. We define an on-site velocity $w=v^n$ as one for which the position-velocity cross-correlation is zero; namely for
\begin{eqnarray}
\langle r^n v^n\rangle & = & 0 \; , \label{eq:cross_cor_wr}
\end{eqnarray}
in accordance with equilibrium statistical mechanics \cite{Langevin_Eq}. Having defined $v^n$ to be the on-site velocity, we can state that
\begin{eqnarray}
\gamma_5 & = & -c_5\gamma_1 \; = \; -\frac{1}{2}c_3\gamma_1 \; , \label{eq:gamma_cond_3}
\end{eqnarray}
since in Eq.~(\ref{eq:recursion_1}) the on-site velocity $v^n$ cannot depend on $\beta^{n+1}$, which is the integrated noise contribution over the interval $t_n< t\le t_{n+1}$. More constraints can be identified if
we insert Eqs.~(\ref{eq:GJF-2GJF}) into Eq.~(\ref{eq:cross_cor_wr}) and obtain
\begin{eqnarray}
\frac{(\gamma_1+\gamma_3)\langle r^nr^{n+1}\rangle+\gamma_2\langle r^nr^n\rangle}{dt}+\gamma_4\frac{\langle r^n\beta^n\rangle}{m} & = & 0 \; , \label{eq:cross_cor_wr2}
\end{eqnarray}
where, for the harmonic oscillator, the needed correlations are given in Eqs.~(\ref{eq:rnbetan}) and (\ref{eq:auto-correlations}). Inserting those correlations into Eq.~(\ref{eq:cross_cor_wr2}) gives the simple on-site relationship
\begin{eqnarray}
\gamma_2 & = & -2\frac{\alpha dt}{m} \, \gamma_4 \; , \label{eq:cross_cor_wr3}
\end{eqnarray}
which applies for any choice of $c_2$. Equations~(\ref{eq:gamma_cond_1}), (\ref{eq:gamma_cond_2}), (\ref{eq:gamma_cond_3}), and (\ref{eq:cross_cor_wr3}) can now be conveniently rewritten
\begin{eqnarray}
\gamma_2 & = & -(1-c_2)\gamma_1  \label{eq:gamma_2_rv_1}\\
\gamma_3 & = & -c_2\gamma_1 \label{eq:gamma_3_rv_1} \\
\gamma_4 & = & -\gamma_5 \; = \; \frac{1}{2}c_3\gamma_1 \; .  \label{eq:gamma_4_rv_1}
\end{eqnarray}
Notice that inserting Eqs.~(\ref{eq:gamma_3_rv_1}) and (\ref{eq:gamma_4_rv_1}) into Eq.~(\ref{eq:condition_G1}) gives
\begin{eqnarray}
\Gamma_1(\frac{\alpha dt}{m}) & = & -c_3^2\gamma_1^2\; , 
\end{eqnarray}
which can only be zero if $\gamma_1=0$. It is, however, clear that $\gamma_1\neq0$.
Thus, we conclude that {\it for any of the statistically correct GJ trajectories from Sec.~\ref{sec:sv_section}, it is not possible to identify a meaningful on-site velocity that produces both correct and time-step-independent kinetic statistics.}

It follows that we must relax a requirement for an on-site velocity. The requirement to relax is Eq.~(\ref{eq:condition_G1}), where $(\Omega_0dt)^2\Gamma_1$ is the error of the kinetic energy if $\Gamma_{-1}=0$ and $\Gamma_0=1$ (see Eq.~(\ref{eq:w2_2})). Thus, $\Gamma_1$ is no longer required to be zero, and we now focus on ensuring that Eq.~(\ref{eq:condition_G0}) is satisfied. Inserting the four conditions, Eqs.~(\ref{eq:gamma_cond_1}), (\ref{eq:gamma_2_rv_1}), (\ref{eq:gamma_3_rv_1}), and (\ref{eq:gamma_4_rv_1}), into Eq.~(\ref{eq:Gamma_0}), we arrive at the condition for Eq.~(\ref{eq:condition_G0}) to be fulfilled
\begin{eqnarray}
\Gamma_0 & = & 4c_1c_3\gamma_1^2 \; = \; 1 \\
\Rightarrow \; \; \gamma_1 & = & \frac{1}{2\sqrt{c_1c_3}} \; . \label{eq:on-site_gamma_1}
\end{eqnarray}
Therefore, while there does not exist an on-site velocity with correct and time-step-independent kinetic statistics for any configurationally correct method, we have shown that, {\it for each GJ St{\o}rmer-Verlet method derived in Sec.~\ref{sec:sv_section}, there exists exactly one on-site velocity, which has correct kinetic statistics in the limit $\Omega_0dt\rightarrow0$, and which has an error no more significant than second order in $\Omega_0dt$}. The on-site velocity can be written in finite difference form
\begin{eqnarray}
v^n & = & \frac{r^{n+1}-(1-c_2)r^n-c_2r^{n-1}}{2\sqrt{c_1c_3}\,dt}+\sqrt{\frac{c_3}{c_1}}\frac{1}{4m}(\beta^n-\beta^{n+1}) \; . \label{eq:GJ_sv_v}
\end{eqnarray}
Using the on-site velocity, the resulting kinetic energy for a harmonic oscillator is
\begin{eqnarray}
\langle E_k\rangle & = &\frac{1}{2}m \langle v^nv^n\rangle \; = \; \frac{k_BT}{2}\left(1-\frac{c_3}{c_1}\frac{\Omega_0^2dt^2}{4}\right) \; .  \label{eq:Ek_GJ}
\end{eqnarray}
Notice that the GJF method (Eqs.~(\ref{eq:gjf_sv_r}) and (\ref{eq:gjf_sv_v}), or Eq.~(\ref{eq:gjf_vv})) is a method of this kind (for $c_2=a$, $c_1=c_3=b$). The GJF method is the single method for which $c_1=c_3$, and the kinetic errors are therefore independent of the friction $\alpha$. However, it is obvious that the error on the on-site velocity is reduced when $|c_3/c_1|$ is small. Thus, in general, methods ${\rm II}$, ${\rm IV}$-${\rm VI}$ yield smaller on-site velocity errors than method ${\rm I}$, which yields smaller errors than those of method ${\rm III}$ (see Fig.~\ref{fig_1}).

Using Eqs.~(\ref{eq:GJ_sv_r}) and (\ref{eq:recursion_1}) with $w^\nu=v^{n+1}$ and with the specific on-site parameters for $\gamma_i$ (Eqs.~(\ref{eq:gamma_2_rv_1})-(\ref{eq:gamma_4_rv_1}) with Eq.~(\ref{eq:on-site_gamma_1})) allow us to write the combined general method in the velocity-Verlet form
\begin{subequations}
\begin{eqnarray}
r^{n+1} & = & r^n+\sqrt{c_1c_3}\,dt\,v^n+\frac{c_3dt^2}{2m}f^n+\frac{c_3dt}{2m}\beta^{n+1}\label{eq:GJ_vv_r}\\
v^{n+1} & = & c_2v^n+\sqrt{\frac{c_3}{c_1}}\frac{dt}{2m}\left(c_2f^n+f^{n+1}\right)+\frac{\sqrt{c_1c_3}}{m}\beta^{n+1} \; . \label{eq:GJ_vv_v}
\end{eqnarray}
\label{eq:GJ_vv}
\end{subequations}
{\bf Equation~(\ref{eq:GJ_vv}) is the VV form of the GJ set of methods, where the GJ trajectory of Eq.~(\ref{eq:GJ_sv_r}) is accompanied by the unique on-site velocity of Eq.~(\ref{eq:GJ_sv_v}).}
For $c_2=a$ this is the GJF method \cite{GJF1}, which is also given in this paper as Eq.~(\ref{eq:gjf_vv}).

%%%%%%%%%%%%%%%%%%%%%%%%%%%%%%%%%%%%%%%%%%%%%%%%%%%%
\subsubsection{Diffusion calculated from on-site velocity autocorrelations}
\label{sec:GK_v}
While not directly relevant for most simulations, it is interesting to compare the configurational Einstein diffusion Eq.~(\ref{eq:gjf_stat_r}) with the comparable kinetic result $D_{GK}$ from the Green-Kubo expression \cite{GreenKubo}
\begin{eqnarray}
D_{GK} & = & \int_0^{\infty}\langle v(t)v(t+s)\rangle_t\,ds \; . \label{eq:D_GK}
\end{eqnarray}
In order to approximate this expression, we need to find the discrete-time velocity-velocity correlation function $\langle v^nv^{n+q}\rangle$. As we wish to make the comparison for a flat surface, $f=0$, we can use the on-site velocity Eq.~(\ref{eq:GJ_vv_v})
\begin{eqnarray}
v^n & = & c_2v^{n-1}+\frac{\sqrt{c_1c_3}}{m}\beta^n \nonumber \\
& = & c_2^nv^0+\frac{\sqrt{c_1c_3}}{m}\sum_{k=0}^{n-1}c_2^{n-k-1}\beta^{k+1}\; , \label{eq:vn}
\end{eqnarray}
from where we for $n\!\rightarrow\!\infty$ find
\begin{eqnarray}
\langle v^nv^{n+q}\rangle & = & c_2^q\,\frac{k_BT}{m } \; . \label{eq:vv_auto-cor}
\end{eqnarray}
As discussed in Ref.~\cite{2GJ}, the Green-Kubo integral Eq.~(\ref{eq:D_GK}) needs to be approximated by a discrete-time Riemann sum, and the three obvious ones are:\\
The right-Riemann sum
\begin{eqnarray}
D_{GK} & \approx & dt\sum_{k=1}^\infty c_2^k\; \frac{k_BT}{m} \; = \; \frac{c_2}{c_3}\frac{k_BT}{\alpha} \; ; \label{eq:GK_v_rR}
\end{eqnarray}
the trapezoidal sum
\begin{eqnarray}
D_{GK} & \approx & dt\left(\frac{1}{2}+\sum_{k=1}^\infty c_2^k\right)\frac{k_BT}{m} \; = \; \frac{c_1}{c_3}\frac{k_BT}{\alpha} \; ; \label{eq:GK_v_tr}
\end{eqnarray}
and the left-Riemann sum
\begin{eqnarray}
D_{GK} & \approx & dt\sum_{k=0}^\infty c_2^k\; \frac{k_BT}{m} \; = \; \frac{1}{c_3}\frac{k_BT}{\alpha} \; , \label{eq:GK_v_lR}
\end{eqnarray}
where $c_2<c_1<1$. We notice that the GJF method (GJ-I in this paper, $c_2=a$, $c_1=c_3=b$) gives the correct $D_{GK}=D_E$ for the trapezoidal approximation, while $c_2=1-\alpha dt/m$ (GJ-III, $c_3=1$) gives the correct Green-Kubo diffusion for the left-Riemann sum. For GJ-V ($c_2=1/(1+\alpha dt/m)=c_3$), we see that the right-Riemann sum gives the correct value $D_{GK}=D_E$. Thus, for $c_2$ in the interval $1\!-\!\alpha dt/m\le c_2\le1/(1\!+\!\alpha dt/m)$, discrete-time approximations to the Green-Kubo integral can conform for any time step with the Einstein diffusion. Notice that $D_{GK}\rightarrow D_E$ for $dt\rightarrow0$ for all methods.

%%%%%%%%%%%%%%%%%%%%%%%%%%%%%%%%%%%%%%%%%%%%%%%%%%%%
\subsection{Half-step velocity $u^{n+\frac{1}{2}}$}
\label{sec:half-step}
As is the case for on-site velocities, a half-step velocity is well defined for $\alpha=0$, since it can be symmetrically expressed by the central difference given in Eq.~(\ref{eq:sv_u_basic}). For the GJF method, a half-step velocity (2GJ) was identified \cite{2GJ} such that kinetic measures for linear systems are correct and time-step-independent; i.e., all the three conditions, Eqs.~(\ref{eq:gamma_cond_1}), (\ref{eq:condition_G0}), and (\ref{eq:condition_G1}), are satisfied. 
We will here generalize and discuss the half-step velocity in light of the new, complete GJ set of stochastic St{\o}rmer-Verlet methods identified in this paper.

We first recall that a complication with obtaining time-step-independent kinetic statistics for on-site velocities is that Eq.~(\ref{eq:condition_G1}) is in conflict with the defined signature of an on-site velocity; namely that the statistics is symmetrically balanced, $\langle v^nr^n\rangle=0$ (see Eq.~(\ref{eq:cross_cor_wr})). This condition is unverifiable for a half-step velocity.

However, a rational definition of a half-step velocity $u^{n+\frac{1}{2}}$ is still needed in order to certify that the velocity interacts statistically correct with the trajectory. We here use the known continuous-time result (see, e.g., chapter 3 in Ref.~\cite{Langevin_Eq}) for a stochastic harmonic oscillator
\begin{eqnarray}
\langle r(t)\dot{r}(t+\tau)\rangle & = & -\sigma\frac{k_BT}{m\Omega_1}\exp(-\frac{\alpha|\tau|}{2m})\sin{\Omega_1|\tau|}\; , \label{eq:rv_cor}
\end{eqnarray}
where $\sigma=1$ for $\tau\ge0$ and $\sigma=-1$ for $\tau<0$, with $\Omega_1^2=\Omega_0^2-(\alpha/2m)^2$. This anti-symmetric expression dictates the
obvious continuous-time result
\begin{eqnarray}
2{\cal S}_{2\tau} & = & \langle r(t)\dot{r}(t+\tau)\rangle+\langle r(t)\dot{r}(t-\tau)\rangle \\
& = & \langle r(t-\tau)\dot{r}(t)\rangle+\langle r(t+\tau)\dot{r}(t)\rangle \; = \; 0 \; .
\end{eqnarray}
Analagous to this, we can define a discrete-time, half-step velocity as one that is symmetrically located between the surrounding positions, $r^n$ and $r^{n+1}$, such that
\begin{eqnarray}
{\cal S}_{dt} & = & 0\, , \label{eq:S=0}
\end{eqnarray}
where
\begin{eqnarray}
2{\cal S}_{dt} & = & \left\langle r^nu^{n+\frac{1}{2}}\right\rangle + \left\langle r^nu^{n-\frac{1}{2}}\right\rangle \; = \; \left\langle r^{n}u^{n+\frac{1}{2}}\right\rangle+\left\langle r^{n+1}u^{n+\frac{1}{2}}\right\rangle \; . \label{eq:cross_corr_ru_dt}
\end{eqnarray}
Evaluating Eq.~(\ref{eq:cross_corr_ru_dt}) with the velocity Eq.~(\ref{eq:GJF-2GJF}), and using the basic condition Eq.~(\ref{eq:gamma_cond_1}), we find
\begin{eqnarray}
2{\cal S}_{dt} & = & \gamma_3\frac{\langle r^{n-1}r^{n+1}\rangle-\langle r^nr^n\rangle}{dt} + \gamma_4\frac{\langle r^n\beta^n\rangle+\langle r^{n+1}\beta^n\rangle}{m}+\gamma_5\frac{\langle r^n\beta^n\rangle}{m} \; . \label{eq:rv_dt}
\end{eqnarray}
Using Eqs.~(\ref{eq:rnp1betan}), (\ref{eq:rnbetan}), and (\ref{eq:auto-correlations}), Eq.~(\ref{eq:rv_dt}) can be written
\begin{eqnarray}
2{\cal S}_{dt} & = & -\gamma_3\frac{k_BT}{\alpha}c_3\frac{\alpha dt}{2m}\left[3+c_2-c_3\Omega_0^2dt^2\right]\nonumber \\
&&+\frac{k_BT}{\alpha}\left(\frac{\alpha dt}{m}\right)^2c_3\left[\gamma_5+2\gamma_4(1+c_1)-\gamma_4c_3\Omega_0^2dt^2\right] \; .
\end{eqnarray}
Satisfying statistical balance of being a true half-step velocity (${\cal S}_{dt}=0$), it follows that $\gamma_3=2\gamma_4\alpha dt/m$. However, this is in conflict with Eq.~(\ref{eq:gamma_cond_2}). The only time-step-independent behavior of ${\cal S}_{dt}$ can be found for $\gamma_3=\gamma_4=\gamma_5=0$, which implies $\gamma_2=-\gamma_1$.
We conclude that {\it all half-step velocities, statistically balanced in the cross-correlation with the configurational coordinate (${\cal S}_{dt}=0$), are described by $\gamma_3=\gamma_4=\gamma_5=0$ $\Rightarrow$ $\gamma_2=-\gamma_1$, and that ${\cal S}_{dt}=0$ for all such velocities.} Notice that the condition $\Gamma_1=0$ from Eq.~(\ref{eq:condition_G1}) is always satisfied if ${\cal S}_{dt}=0$. We determine $\gamma_1$ by the condition Eq.~(\ref{eq:condition_G0}), which in this case reads
\begin{eqnarray}
\Gamma_0 & = & \gamma_1^2c_3 \; = \; 1 \label{eq:gam_G0_GJ} \\
\Rightarrow \; \; \gamma_1 & = & \frac{1}{\sqrt{c_3}} \; .\label{eq:gam1_GJ}
\end{eqnarray}
We then have the statistically correct, time-step-independent half-step velocity $u^{n+\frac{1}{2}}$ given by
\begin{eqnarray}
u^{n+\frac{1}{2}} & = & \frac{r^{n+1}-r^n}{\sqrt{c_3}\,dt}  \; .\label{eq:GJ_u}
\end{eqnarray}
We conclude that {\it for each statistically correct stochastic St{\o}rmer-Verlet method derived in Sec.~\ref{sec:sv_section}, there exists exactly one half-step velocity, which has correct kinetic statistics independent of the time step.}
Notice that for the GJF method, where $c_2=a$ and $c_3=b$, this is the 2GJ half-step velocity from Eq.~(\ref{eq:2gj}) that was identified in Ref.~\cite{2GJ}.

Equations~(\ref{eq:GJ_sv_r}) and (\ref{eq:GJ_u}) allow us to write the combined general method in the leap-frog form
\begin{subequations}
\begin{eqnarray}
u^{n+\frac{1}{2}} & = & c_2u^{n-\frac{1}{2}}+\sqrt{c_3}\left[\frac{dt}{m}f^n+\frac{1}{2m}(\beta^n+\beta^{n+1})\right] \label{eq:GJ_lf_u}\\
r^{n+1} & = & r^n + \sqrt{c_3}\,dt\, u^{n+\frac{1}{2}} \; . \label{eq:GJ_lf_r}
\end{eqnarray}
\label{eq:GJ_lf}
\end{subequations}
{\bf Equation~(\ref{eq:GJ_lf}) is the LF form of the GJ set of methods, where the GJ trajectory of Eq.~(\ref{eq:GJ_sv_r}) is accompanied by the unique half-step velocity of Eq.~(\ref{eq:GJ_u}).} For $c_2=a$ this form is the GJF-2GJ method of Ref.~\cite{2GJ}.

%%%%%%%%%%%%%%%%%%%%%%%%%%%%%%%%%%%%%%%%%%%%%%%%%%%%
\subsubsection{Diffusion calculated from half-step velocity autocorrelations}
\label{sec:GK_u}
We here calculate Green-Kubo approximations to diffusion on a flat surface using the identified half-step velocity. From Eq.~(\ref{eq:GJ_lf_u}), for $f=0$, we write the half-step velocity as
\begin{eqnarray}
u^{n+\frac{1}{2}} & = & c_2u^{n-\frac{1}{2}}+\frac{\sqrt{c_3}}{2m}(\beta^n+\beta^{n+1}) \\
& = & c_2^n\,u^\frac{1}{2} + \frac{\sqrt{c_3}}{2m}\left[c_2^{n-1}\beta^1+\beta^{n+1}+\frac{2c_1}{c_2}\sum_{k=1}^{n-1}c_2^{n-k}\beta^{k+1}\right] \; ,
\end{eqnarray}
where we have used Eq.~(\ref{eq:vn}).  This expression leads us to the velocity-velocity autocorrelation for $n\rightarrow\infty$
\begin{eqnarray}
\langle u^{n+\frac{1}{2}}u^{n+\frac{1}{2}+q}\rangle & = & \frac{k_BT}{m}\times\left\{\begin{array}{ccc} 1 & , & q=0 \\ c_1c_2^{q-1} & , & q\ge1\end{array}\right. \; . \label{eq:uu_auto-cor}
\end{eqnarray}
The three characteristic discrete-time approximations to the Green-Kubo integral Eq.~(\ref{eq:D_GK}) are:\\
The right-Riemann sum
\begin{subequations}
\begin{eqnarray}
D_{GK} & \approx & dt\frac{c_1}{c_2}\sum_{k=1}^\infty c_2^k\; \frac{k_BT}{m} \; = \; \frac{c_1}{c_3}\frac{k_BT}{\alpha} \; ; \label{eq:GK_u_rR}
\end{eqnarray}
the trapezoidal sum
\begin{eqnarray}
D_{GK} & \approx & dt\left(\frac{1}{2}+\frac{c_1}{c_2}\sum_{k=1}^\infty c_2^k\right)\;\frac{k_BT}{m} \; = \; \frac{1}{c_3}\frac{k_BT}{\alpha}\; ; \label{eq:GK_u_tr}
\end{eqnarray}
and the left-Riemann sum
\begin{eqnarray}
D_{GK} & \approx & dt\left(1+\frac{c_1}{c_2}\sum_{k=1}^\infty c_2^k\right)\; \frac{k_BT}{m} \; = \; (\frac{1}{c_3}+\frac{\alpha dt}{2m})\frac{k_BT}{\alpha} \; ,\label{eq:GK_u_lR}
\end{eqnarray}
\end{subequations}
where $c_2<c_1<1$. We here highlight the GJF-2GJ method (GJ-I in this paper, $c_2=a$, $c_1=c_3=b$), which gives the correct $D_{GK}=D_E$ for the right-Riemann approximation, while $c_2=1-\alpha dt/m$ (GJ-III, $c_3=1$) gives the correct Green-Kubo diffusion for the trapezoidal sum. For $c_2=(2-1/a)$, we see that the left-Riemann sum gives the correct value $D_{GK}=D_E$. Notice that $D_{GK}\rightarrow D_E$ for $dt\rightarrow0$ for all methods.

%%%%%%%%%%%%%%%%%%%%%%%%%%%%%%%%%%%%%%%%%%%%%%%%%%%%
\section{Compact form of the GJ algorithms}
\label{sec:combo}

We can combine the statistically correct on-site Eq.~(\ref{eq:GJ_sv_v}) and half-step Eq.~(\ref{eq:GJ_u}) velocities with the GJ trajectory, Eq.~(\ref{eq:GJ_sv_r}), in the compact form
\begin{subequations}
\begin{eqnarray}
u^{n+\frac{1}{2}} & = & \sqrt{c_1}\,v^n+\frac{\sqrt{c_3}\,dt}{2m}f^n+\frac{\sqrt{c_3}}{2m}\beta^{n+1}\label{eq:compact_u}\\
r^{n+1} & = & r^n+\sqrt{c_3}\,dt\, u^{n+\frac{1}{2}} \label{eq:compact_r}\\
v^{n+1} & = & \frac{c_2}{\sqrt{c_1}}u^{n+\frac{1}{2}}+\sqrt{\frac{c_3}{c_1}}\frac{dt}{2m}f^{n+1}+\sqrt{\frac{c_3}{c_1}}\frac{1}{2m}\beta^{n+1} \; , \label{eq:compact_v}
\end{eqnarray}
\label{eq:compact_all}
\end{subequations}
with the noise given by Eq.~(\ref{eq:noise_dt}). 
{\bf Equation~(\ref{eq:compact_all}) is the GJ set of methods, where the GJ trajectory of Eq.~(\ref{eq:GJ_sv_r}) is accompanied by the unique on-site velocity of Eq.~(\ref{eq:GJ_sv_v}) as well as the unique half-step velocity of Eq.~(\ref{eq:GJ_u}).} For linear systems, this GJ set of methods provides correct, time-step-independent statistics for both kinetic and configurational measures when using the variables $r^n$ and $u^{n+\frac{1}{2}}$, the latter being the statistically balanced, true half-step velocity. The equations can equally well be listed in St{\o}rmer-Verlet form [Eqs.~(\ref{eq:GJ_sv_r}) and (\ref{eq:GJ_sv_v})], in velocity-Verlet form [Eq.~(\ref{eq:GJ_vv})], and in leap-frog form [Eq.~(\ref{eq:GJ_lf})]. The variable $v^n$ is the statistically balanced on-site velocity, which has quadratic leading statistical error $\propto(\Omega_0dt)^2$ in the kinetic energy measure. The complete set of methods is given by a single functional parameter, e.g., $c_2$ ($|c_2|\le1$), which is the velocity attenuation factor over one time step, and the relationships between parameters can be found in Table~\ref{tab:sv}. For $c_2=a$, Eq.~(\ref{eq:compact_all}) coincides with the comparable set of equations given for the GJF-2GJ method from Ref.~\cite{2GJ2}. We notice that method IV, mentioned in Sec.~\ref{sec:sv_highlight}, makes all prefactors in Eq.~(\ref{eq:compact_v}) the same.

While any method in this set of methods will behave statistically correct in linear equilibrium conditions with the trajectories $r^n$ and half-step velocities $u^{n+\frac{1}{2}}$ for any time step within the stability limit, we notice that the drift velocity for $f^n=f={\rm const}$ can be incorrectly measured by both on-site and half-step velocities, even if the drift is correctly represented by the trajectory Eq.~(\ref{eq:r_drift}). The values for the two types of velocities are found from Eqs.~(\ref{eq:GJ_sv_v}) and (\ref{eq:GJ_u}) to be
\begin{eqnarray}
\langle v^{n} \rangle & = &\sqrt{\frac{c_1}{c_3}} \, \frac{f}{\alpha}\; , \label{eq:v_drift}
\end{eqnarray}
which is correct only for $c_2=a$ (the GJF method, where $c_3=c_1=b$), and
\begin{eqnarray}
\langle u^{n+\frac{1}{2}} \rangle & = &\frac{1}{\sqrt{c_3}} \, \frac{f}{\alpha} \; ,  \label{eq:u_drift}
\end{eqnarray}
which is correct only for $c_2=1-\alpha dt/m$ (GJ-III, where $c_3=1$). Clearly, it is desirable if a method can produce both correct statistics and correct drift (ballistic) properties in a single velocity definition. Embedded in a single velocity, this can be accomplished only through the half-step velocity $u^{n+\frac{1}{2}}$ for $c_2=1-\frac{\alpha dt}{m}$ (GJ-III).

%%%%%%%%%%%%%%%%%%%%%%%%%%%%%%%%%%%%%%%%%%%%%%%%%%%%
\section{Other, neither on-site nor half-step velocities}
\label{sec:other_velocities}
One may decide to relax the condition on statistical symmetry of a velocity definition; i.e., consider a choice that places the velocity neither on-site (Eq.~(\ref{eq:cross_cor_wr})) nor at half-step (Eq.~(\ref{eq:S=0})). There may be numerous reasons for attempting this, one being that, given the seemingly poor stability properties for large damping of method GJ-III above, it may be deemed more important to obtain correct drift velocity with good stability than it is to have correct statistical symmetry. While this is statistically questionable, we will here generalize Ref.~\cite{GJF-2} and evaluate the two-point velocity expression that is not necessarily at half-step. Similarly, we will briefly analyze three-point velocity expressions that are not necessarily on-site.

%%%%%%%%%%%%%%%%%%%%%%%%%%%%%%%%%%%%%%%%%%%%%%%%%%%%
\subsection{Two-point velocity $u^{n+\frac{1}{2}+\varepsilon}$}
We define a two-point velocity as one with $\gamma_3=0$. The two-point velocity is attractive since it automatically satisfies the condition Eq.~(\ref{eq:condition_G1}) with $\gamma_3=\gamma_4=0$. Since $\gamma_1+\gamma_2+\gamma_3=0$ (see Eq.~(\ref{eq:gamma_cond_1})), we have in this case that $\gamma_2=-\gamma_1$, further simplifying the evaluation of the condition Eq.~(\ref{eq:condition_G0}), which then reads
\begin{eqnarray}
\Gamma_{0} & = & c_3\gamma_1^2+2\frac{\alpha dt}{m}(c_3\gamma_1\gamma_5+\gamma_5^2) \; = \; 1 \; . \label{eq:cond_two-point}
\end{eqnarray}
This yields the relationship
\begin{eqnarray}
\gamma_5 & = & -\frac{c_3}{2}\gamma_1\pm\frac{1}{2}\sqrt{\frac{2c_3}{1-c_2}(1-c_1c_3\gamma_1^2)} \; , \label{eq:link_15}
\end{eqnarray}
where $c_1c_3\gamma_1^2\le1$, and where we should choose the solution with "+" (see below). Thus, for each GJ method (given by, e.g., a choice of $c_2$) we have a set of two-point velocities of the form
\begin{eqnarray}
u_{\gamma_1}^{n+\frac{1}{2}+\varepsilon} & = & \gamma_1\frac{r^{n+1}-r^n}{dt}+\gamma_5\frac{\beta^{n+1}}{m} \; , \label{eq:two-point_u}
\end{eqnarray}
where $\gamma_5$ is linked to $\gamma_1$ through Eq.~(\ref{eq:link_15}), and where it is reasonable to choose $0<\gamma_1\sqrt{c_1c_3}\le1$. For the specific choice $c_2=a$, this is the set of two-point velocities given in Ref.~\cite{GJF-2}. For $\gamma_5=0$, which is accomplished for $\gamma_1\sqrt{c_3}=1$, we recover the half-step velocity given in Eq.~(\ref{eq:GJ_u}).

As mentioned above, one reason for relaxing the requirement for statistical symmetry can be that the two-point velocity, which is statistically correct in its average kinetic energy, can also yield correct drift (ballistic) velocity for any choice of GJ method, $c_2$. This is obviously accomplished for $\gamma_1=1$. Specifically,
\begin{eqnarray}
u_1^{n+\frac{1}{2}+\varepsilon} & = & \frac{r^{n+1}-r^n}{dt}+\gamma_5\frac{\beta^{n+1}}{m} 
\end{eqnarray}
is a two-point velocity with correct drift and ballistic transport. The associated value of $\gamma_5$ is easily found from Eq.~(\ref{eq:link_15}). We notice that for $\gamma_1=1$, Eq.~(\ref{eq:link_15}) shows that $\gamma_5\ge0$ if the "+" sign is chosen.

Since the two-point velocity is not necessarily located at the half-step between positions, we will estimate the relative temporal location $\varepsilon$ of the velocity by evaluating the cross-correlations given in Eqs.~(\ref{eq:rv_cor}) and (\ref{eq:rv_dt}). For the two-point velocity in Eq.~(\ref{eq:two-point_u}), we have
\begin{eqnarray}
2{\cal S}_{dt} & = & \gamma_5\left(\frac{\alpha dt}{m}\right)^2c_3\frac{k_BT}{\alpha}\; , \label{eq:S_dt_g5}
\end{eqnarray}
which is clearly zero only for $\gamma_5=0$. In order to determine which fractional time step $\varepsilon$ this corresponds to, we use Eq.~(\ref{eq:rv_cor}) and evaluate the continuous-time equivalent of $2{\cal S}_{dt}$
\begin{eqnarray}
2\tilde{\cal S}_{dt} & = & \left\langle r(t)\dot{r}(t+\frac{dt}{2}+\varepsilon dt)\right\rangle+\left\langle r(t)\dot{r}(t-\frac{dt}{2}+\varepsilon dt)\right\rangle \\
& = & 2\frac{k_BT}{m\Omega_1}e^{-\frac{\alpha dt}{4m}}\left[\sinh\frac{\alpha dt}{2m}\varepsilon\,\sin\frac{\Omega_1dt}{2}\,\cos\Omega_1dt\varepsilon-\cosh\frac{\alpha dt}{2m}\varepsilon\,\cos\frac{\Omega_1dt}{2}\,\sin\Omega_1dt\varepsilon\right] \; , %\nonumber \\
\end{eqnarray}
where we assume that $-1\le2\varepsilon\le1$.
If we compare Eq.~(\ref{eq:S_dt_g5}) to $2\tilde{\cal S}_{dt}$ for $\Omega_1dt\ll1$ and $\alpha dt\ll2m$ we get
\begin{eqnarray}
2{\cal S}_{dt} & \approx & -2\frac{k_BT}{m}dt\,\varepsilon \; \; \; \Rightarrow \\
\varepsilon & \approx & -\frac{\alpha dt}{2m}\,c_3\gamma_5 \; , 
\end{eqnarray}
which is also the result found for overdamped systems. The small-$dt$ approximation indicates that the deviation from half-step is proportional to $\gamma_5$, and that a positive $\gamma_5$ puts the two-point velocity at a time behind the half-step. Thus, the best sign to choose in Eq.~(\ref{eq:link_15}) is "+", since that will minimize the magnitude of $\gamma_5$, given that we choose $\gamma_1>0$.

The specific compact form using the general two-point velocity $u_{\gamma_1}^{n+\frac{1}{2}+\varepsilon}$ along with the on-site velocity $v^n$ is
\begin{subequations}
\begin{eqnarray}
u_{\gamma_1}^{n+\frac{1}{2}+\varepsilon} & = & \gamma_1\sqrt{c_1c_3}\,v^n+\gamma_1\frac{c_3\,dt}{2m}f^n+\left(\frac{\gamma_5}{m}+\gamma_1\frac{c_3}{2m}\right)\beta^{n+1}\label{eq:tp_gen_compact_u}\\
r^{n+1} & = & r^n+\frac{dt}{\gamma_1}\left[u_{\gamma_1}^{n+\frac{1}{2}+\varepsilon} - \frac{\gamma_5}{m}\,\beta^{n+1}\right]\label{eq:tp_gen_compact_r}\\
v^{n+1} & = & \frac{c_2}{\gamma_1\sqrt{c_1c_3}}u_{\gamma_1}^{n+\frac{1}{2}+\varepsilon}+\sqrt{\frac{c_3}{c_1}}\frac{dt}{2m}f^{n+1}-\frac{1}{\gamma_1\sqrt{c_1c_3}}\left(\frac{\gamma_5}{m}c_2-\gamma_1\frac{c_3}{2m}\right)\beta^{n+1} \; , %\nonumber \\ 
\label{eq:tp_gen_compact_v}
\end{eqnarray}
\end{subequations}
with the noise given by Eq.~(\ref{eq:noise_dt}), $\gamma_5$ is given by Eq.~(\ref{eq:link_15}), and $c_1c_3\gamma_1^2\le1$. The correct drift velocity is found when using $u_{\gamma_1}^{n+\frac{1}{2}+\varepsilon}$ for $\gamma_1=1$, and the true half-step velocity $u^{n+\frac{1}{2}}$ is found for $\gamma_1=1/\sqrt{c_3}$ ($\gamma_5=0$).

An example of a set of such velocities was recently proposed \cite{GJF-2} for the GJF method ($c_2=a$) in the form of a two-point velocity with $\gamma_3=\gamma_4=0$. One special case of this set is the GJF-2GJ velocity, which is the true half-step velocity for the GJF (GJ-I) method. The rest of the
two-point velocities in this set are, as outlined above, not statistically located at the half step, and located differently for different values of $\alpha dt/2m$. Thus, while these kinds of velocities are designed to give correct kinetic measures, drift and diffusion transport for $\gamma_1=1$, they are not necessarily well coordinated with the configurational measures obtained from $r^n$, and may therefore cause complications for statistical quantities that, e.g., mix kinetic with configurational sampling.

%%%%%%%%%%%%%%%%%%%%%%%%%%%%%%%%%%%%%%%%%%%%%%%%%%%%
\subsection{Three-point velocity $v^{n+\varepsilon}$}
We have determined in Sec.~\ref{sec:on-site} that a statistically correct and time-step-independent on-site velocity cannot be sensibly defined. We have also identified the three-point on-site velocity that is correct to second order in $\Omega_0dt$. Relaxing the on-site condition given in Eq.~(\ref{eq:cross_cor_wr}), one could seek to improve the three-point velocity by making the resulting velocity time-step independent in its statistical response. Thus, one would seek to satisfy the condition $\Gamma_1=0$ with Eq.~(\ref{eq:Gamma_1}). However, since $\gamma_1\neq0$ and since Eq.~(\ref{eq:gamma_cond_2}) must be satisfied for all finite difference velocities of the kind under investigation, we find that Eqs.~(\ref{eq:gamma_cond_2}) and (\ref{eq:condition_G1}) are always in conflict for a three-point velocity, where both $\gamma_1$ and $\gamma_3$ are non-zero. Thus, relaxing the on-site condition for a three-point velocity does not lead to an improvement in the statistical response over the already defined on-site velocity from Sec.~\ref{sec:on-site}. We submit that {\it the on-site velocity is the statistically best possible three-point velocity that can be associated with a GJ trajectory.}

An interesting exercise is to apply the basic three-point, central difference velocity Eq.~(\ref{eq:sv_v_basic}) to a statistically correct GJ trajectory. This velocity has the parameters $\gamma_1=-\gamma_3=\frac{1}{2}$ and $\gamma_2=\gamma_4=\gamma_5=0$. Inserting these values into the velocity conditions gives the following assessment: Neither Eq.~(\ref{eq:gamma_cond_2}) nor (\ref{eq:gamma_cond_3}) are satisfied, so the velocity is not on-site  for $\alpha>0$. Equation~(\ref{eq:gamma_cond_1}) is satisfied, so the velocity does not diverge for $dt\rightarrow0$. Equations~(\ref{eq:condition_G0}) and (\ref{eq:condition_G1}) are not satisfied, so the velocity is not statistically time-step-independent. The resulting average kinetic energy is found from Eq.~(\ref{eq:w2_2}) to be
\begin{eqnarray}
\langle E_k\rangle & = & \frac{1}{2}k_BT \, c_3\left(\frac{1+c_1}{2}-c_3\frac{\Omega_0^2dt^2}{4}\right) \; . \label{eq:Ekin_w}
\end{eqnarray}
This expression has the correct limit $\langle E_k\rangle\rightarrow\frac{1}{2}k_BT$ for $dt\rightarrow0$. However, it has dramatic first order errors in $\alpha dt/m$ due to the coefficients $c_1$ and $c_3$. It follows from the relationship $v^{n+\frac{1}{2}}=\sqrt{c_3}u^{n+\frac{1}{2}}$ (see Eqs.~(\ref{eq:sv_u_basic}) and (\ref{eq:GJ_u})) that even the symmetric average $w^n=\frac{1}{2}(u^{n+\frac{1}{2}}+u^{n-\frac{1}{2}})$ of statistically correct half-step velocities does not yield a statistically correct on-site velocity. The average is not on-site, according to Eqs.~(\ref{eq:gamma_cond_2}) and  (\ref{eq:gamma_cond_3}), and the kinetic energy measured from $w^n$ yields $c_3^{-1}$ of the result in Eq.~(\ref{eq:Ekin_w}), which is not correct for $dt>0$.

We conclude this section by reemphasizing that the on-site velocity defined in Sec.~\ref{sec:on-site} is the optimal three-point velocity, and that statistically correct kinetic information must be extracted from the half-step velocity.

%%%%%%%%%%%%%%%%%%%%%%%%%%%%%%%%%%%%%%%%%%%%%%%%%%%%
\section{Numerical simulations}
\label{sec:numerical_simulations}
In order to confirm the applicability and features of the methods derived in this paper, we show results of model simulations of GJ methods I-VI for both simple, nonlinear, one-dimensional systems and complex Molecular Dynamics. Figure~\ref{fig_1} indicates three different kinds of behavior of the exemplified methods as a function of $\alpha dt/m$. GJ-II, IV-VI all have $c_2\rightarrow0$ for $\alpha dt/m\rightarrow\infty$, implying that these methods will smoothly become GJ-0 in the overdamped limit. GJ-II enters into an unphysical dynamical regime for $c_2<0$ when $\alpha dt/m>2$, but the method always satisfies $|c_2|<1$. Common for Methods I, II, IV-VI is that their stability properties are not restricted by an increase in the damping $\alpha$. In contrast, GJ-III, which is the method that satisfies most statistical properties, becomes unstable for increasing $\alpha$. Particular expectations from the above analysis include that the kinetic properties derived from the on-site velocity should be better when $|c_3/c_1|$ is minimized (see Eq.~(\ref{eq:Ek_GJ})). Thus, we expect that increasing $\alpha$ will result in on-site kinetic results worsening for GJ-III, not influencing GJ-I, and improving the rest. This feature is apparent in all the examples shown below, and especially visible for the simulations with large damping. We recall that the drift velocity will be different from method to method, since the measured drift depends on $c_1$ and $c_3$ (see Eqs.~(\ref{eq:v_drift}) and (\ref{eq:u_drift})). The overall expectation is, however, that all methods should behave similarly for as long as they are applied within their stability ranges. 
For all numerical simulation results, we have used the compact form of the GJ methods given in Eq.~(\ref{eq:compact_all}).

%%%%%%%%%%%%%%%%%%%%%%%%%%%%%%%%%%%%%%%%%%%%%%%%%%%%
\subsection{One-dimensional model systems}
\label{sec:one-dimensional}
We will here study the statistical behavior of the algorithms in simple, nonlinear systems with one degree of freedom. The equation of motion is given by Eq.~(\ref{eq:Langevin}) for which we study two different kinds of potential energy surfaces, $E_1(r)$ and $E_2(r)$, given by
\begin{eqnarray}
\frac{E_{p1}(r)}{E_0} & = & \left(1-\frac{|r|}{r_0}\right)^{-12}-2\left(1-\frac{|r|}{r_0}\right)^{-6}+1 \; \;, \; \;  |r|<r_0 \label{eq:E_p1}\\
\frac{E_{p2}(r)}{E_0} & = & \frac{1}{2}\frac{\kappa r_0^2}{E_0}\,\left(\frac{r}{r_0}\right)^2+1-\cos(\frac{r}{r_0}-\theta) \; , \label{eq:E_p2}
\end{eqnarray}
where $E_0$ is a characteristic energy, $r_0$ is a characteristic distance, and $\theta$ is a constant angle.
These potentials have previously been used to validate the behavior of the GJF method (see, e.g., Refs.~\cite{GJF4,2GJ}). While the potentials have been chosen for reasons of illustration and not physical importance, we note that $E_{p1}$ is a convex potential, consisting of two symmetrically opposing Lennard-Jones cores \cite{AllenTildesley}, relevant for Molecular Dynamics, and that $E_{p2}$ can represent, e.g., the energy surface in a superconducting quantum interference device (SQUID) \cite{Barone}. Both potentials are highly nonlinear, which is essential for the simulations, since all linear expectations are resolved by the analysis in the previous sections. See Figs.~\ref{fig_6}b and \ref{fig_10}b for visual displays of the potentials. For each potential $E_p$ we find the force
\begin{eqnarray}
f & = & -\frac{\partial E_p}{\partial r} \; , 
\end{eqnarray}
and we define the characteristic time scale from the inverse of the frequency $\omega_0=\sqrt{E_0/m}/r_0$. Notice that the natural frequencies of the two systems are different from $\omega_0$. The most relevant natural frequency $\Omega_0$ is found from the curvature of the potential in its deepest ground state, where the potential curvature is largest. Thus, the natural frequency is given by $\Omega_0^2=E_p^{\prime\prime}(r^*)/m$, where $r^*$ is a fixed point. In case of $E_{p1}$, the fixed point $r^*=0$ yields $\Omega_0^2=E_{p1}^{\prime\prime}(0)/m=72\omega_0^2$. For $E_{p2}$, $r^*\approx\theta$ for $\theta\in(-\pi,\pi]$, yielding $\Omega_0^2\approx\frac{\kappa}{m}+\omega_0^2$ (for $\frac{\kappa}{m}\ll\omega_0^2$).

\begin{figure}[t]
\centering
\scalebox{0.8}{\centering \includegraphics[trim={1.5cm 7.0cm 1.5cm 2.0cm},clip]{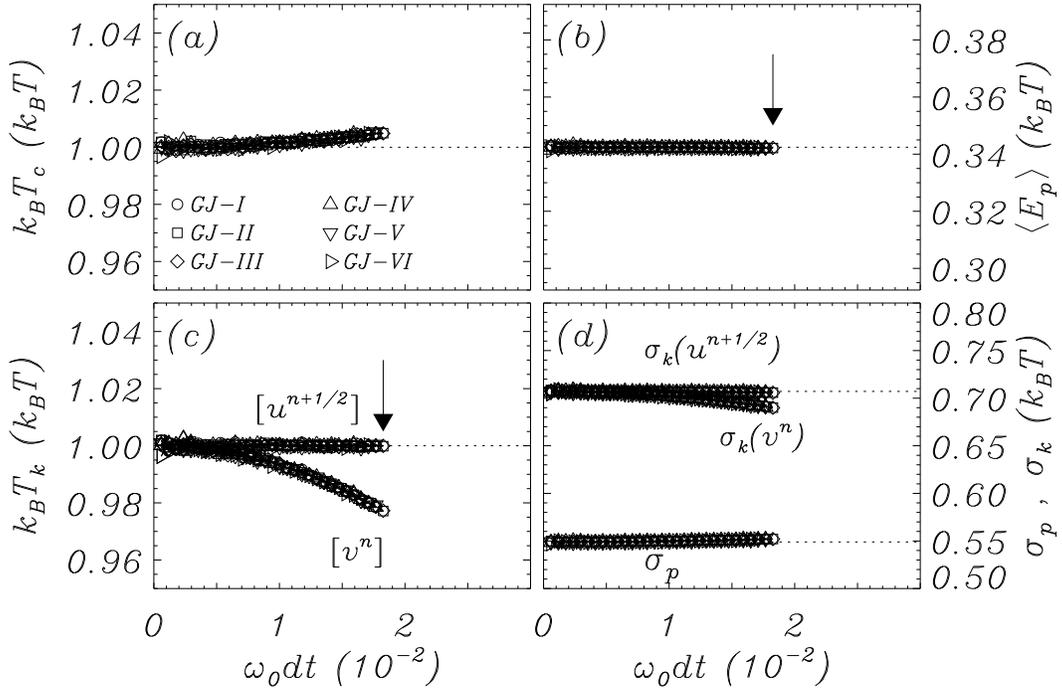}}
%\vspace{-0.50 cm}
\caption{GJ simulations of the one-dimensional oscillator described by $E_{p1}(r)$ in Eq.~(\ref{eq:E_p1}) with $\alpha/m\omega_0=0.1$ and $k_BT=E_0$. Statistics for each data point is acquired from simulations of 10$^{10}$ time steps. Data shown for methods
highlighted in Sec.~\ref{sec:sv_highlight} with on-site velocity $v^n$ and half-step velocity $u^{n+\frac{1}{2}}$ given in Eqs.~(\ref{eq:GJ_sv_v}) and (\ref{eq:GJ_u}). The entire stability range in $\omega_0dt$ is shown for the simulated temperature. Actual stability limit is considerably lower than linearly }
\label{fig_3}
\end{figure}

\begin{figure}[t]
\centering
\scalebox{0.8}{\centering \includegraphics[trim={1.5cm 7.0cm 1.5cm 2.0cm},clip]{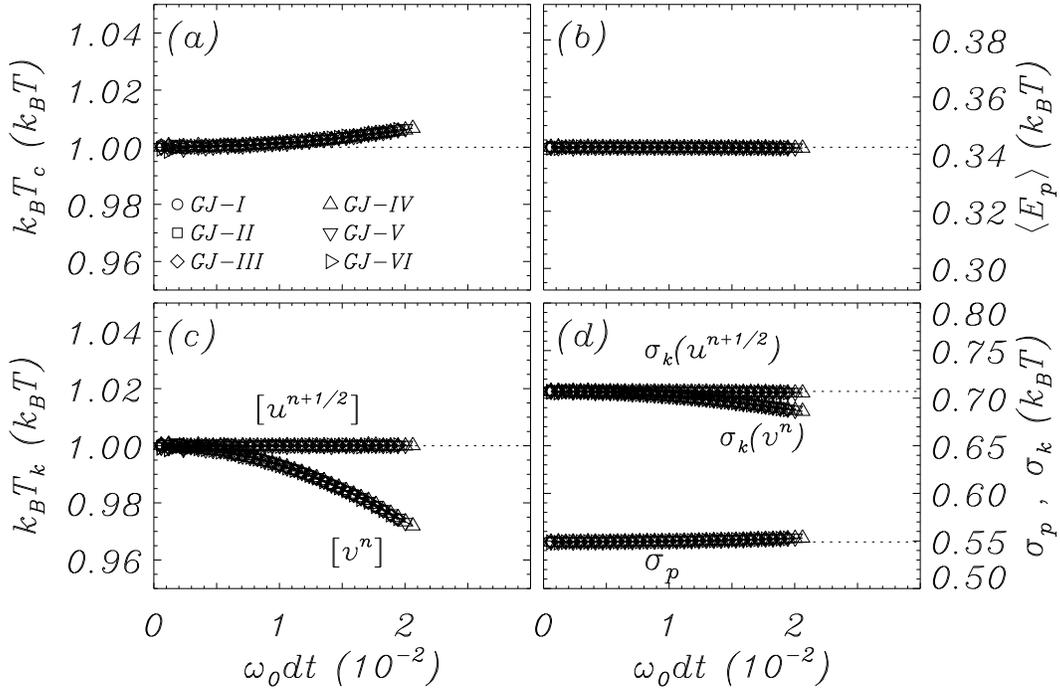}}
%\vspace{-0.50 cm}
\caption{GJ simulations of the one-dimensional oscillator described by $E_{p1}(r)$ in Eq.~(\ref{eq:E_p1}) with $\alpha/m\omega_0=1$ and $k_BT=E_0$.
Other simulation and figure information are given in the caption of Fig.~\ref{fig_3}.
Statistics for each data point is acquired from simulations of 10$^{10}$ time steps. Data shown for methods ${\rm I}$ ($\circ$), ${\rm II}$ ($\Box$) and ${\rm III}$ ($\Diamond$), highlighted in Sec.~\ref{sec:sv_highlight} with on-site velocity $v^n$ and half-step velocity $u^{n+\frac{1}{2}}$ given in Eqs.~(\ref{eq:GJ_sv_v}) and (\ref{eq:GJ_u}). The entire stability range in $\omega_0dt$ is shown for the simulated temperature. Actual stability limit is considerably lower than linearly predicted $\Omega_0dt=\sqrt{72}\omega_0dt<2$ due to strong convex nonlinearity of the potential at the simulated temperature. See also Figs.~\ref{fig_1} and \ref{fig_2} for method coefficients and linear stability. Actual equations simulated are in the form of Eqs.~(\ref{eq:compact_all}). (a) Configurational temperature $T_c$ given by Eq.~(\ref{eq:Tc}). Horizontal dotted line is $k_BT_c=k_BT$. (b) Average potential energy given by the sampling of $E_{p1}$ in Eq.~(\ref{eq:E_p1}). Horizontal dotted line is given by the expected value from Eq.~(\ref{eq:moments}). (c) Kinetic temperature $T_k$ given by Eq.~(\ref{eq:Tk}). Kinetic temperatures are given using both on-site and half-step velocities. Horizontal dotted line is $k_BT_k=k_BT$. (d) Fluctuations, $\sigma_k$ and $\sigma_p$, of kinetic and potential energies. Kinetic fluctuations are shown for both on-site and half-step velocities. Horizontal dotted lines for $\sigma_k$ represent the expectation $\sigma_k=k_BT/\sqrt{2}$. Horizontal dotted line for $\sigma_p$ represents the expected result given by Eq.~(\ref{eq:moments}).
}
\label{fig_4}
\end{figure}

\begin{figure}[t]
\centering
\scalebox{0.8}{\centering \includegraphics[trim={1.5cm 7.0cm 1.5cm 2.0cm},clip]{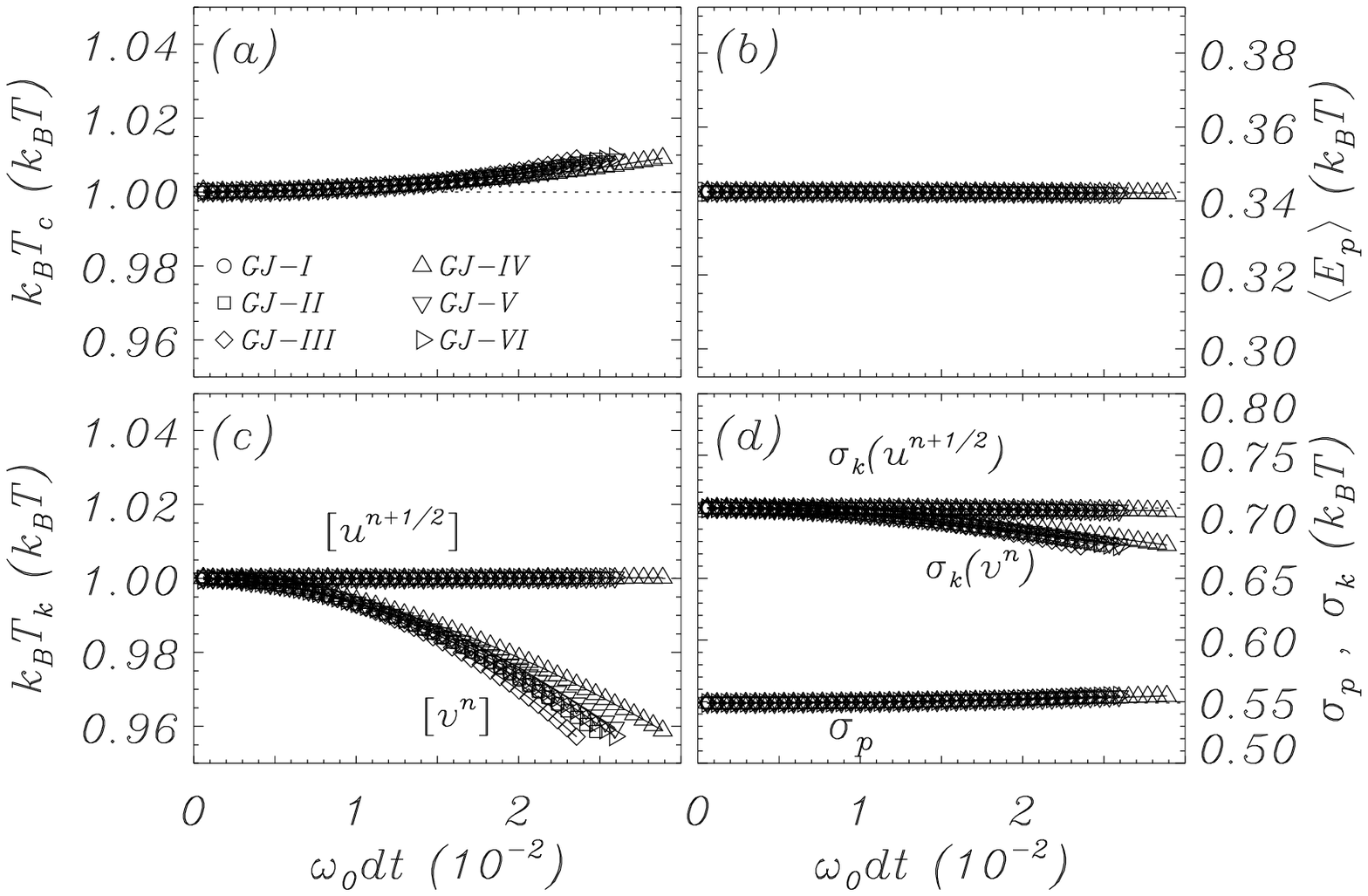}}
%\vspace{-0.50 cm}
\caption{GJ simulations of the one-dimensional oscillator described by $E_{p1}(r)$ in Eq.~(\ref{eq:E_p1}) with $\alpha/m\omega_0=10$ and $k_BT=E_0$. 
Other simulation and figure information are given in the caption of Fig.~\ref{fig_3}.
}
\label{fig_5}
\end{figure}

\begin{figure}[t]
\centering
\scalebox{0.8}{\centering \includegraphics[trim={1.5cm 4.0cm 1.5cm 5.0cm},clip]{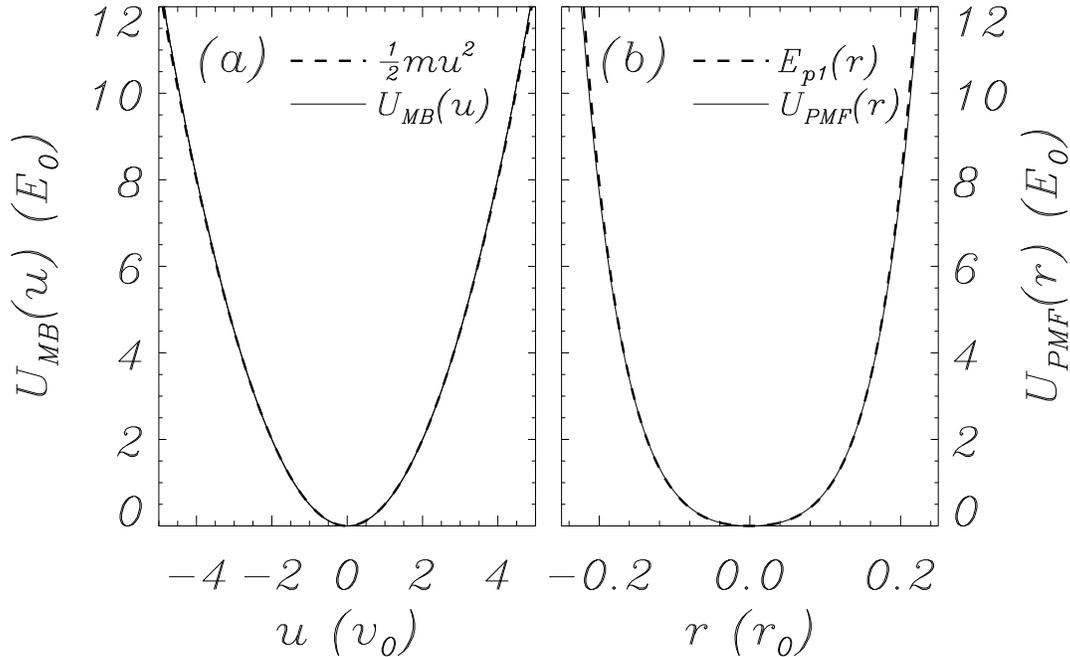}}
%\vspace{-0.50 cm}
\caption{GJ simulation of effective potentials $U_{MB}(u)$ and $U_{PMF}(r)$ in configuration $r^n$ and velocity $u^{n+\frac{1}{2}}$ using GJ method III, highlighted in Sec.~\ref{sec:sv_highlight}, for $E_{p1}(r)$ in Eq.~(\ref{eq:E_p1}). Parameters are $\alpha/m\omega_0=0.1$, $k_BT=E_0$, and $\omega_0dt\approx0.01827$, which is at the edge of the stability limit, as indicated by arrows in Fig.~\ref{fig_3}bc. Actual stability limit is considerably lower than linearly predicted $\Omega_0dt=\sqrt{72}\omega_0dt<2$ due to strong convex nonlinearity of the potential at the simulated temperature. Statistics is acquired from simulations of 10$^{10}$ time steps.  See also Figs.~\ref{fig_1} and \ref{fig_2} for method coefficients and linear stability. Actual equations simulated are in the form of Eq.~(\ref{eq:compact_all}). (a) Effective kinetic potential as a function of the half-step velocity. Solid curve is $U_{MB}(u)$ as measured from the distribution function $\rho_k$ given in Eq.~(\ref{eq:U_MB}); dashed curve is the continuous time expectation. (b) Effective potentials of mean force as a function of the coordinate $r^n$. Solid curve is $U_{PMF}(r)$ as measured from the distribution function $\rho_c$ given in Eq.~(\ref{eq:U_PMF}); dashed curve is the continuous time expectation from Eq.~(\ref{eq:E_p1}).
}
\label{fig_6}
\end{figure}

\begin{figure}[t]
\centering
\scalebox{0.8}{\centering \includegraphics[trim={1.5cm 7.0cm 1.5cm 2.0cm},clip]{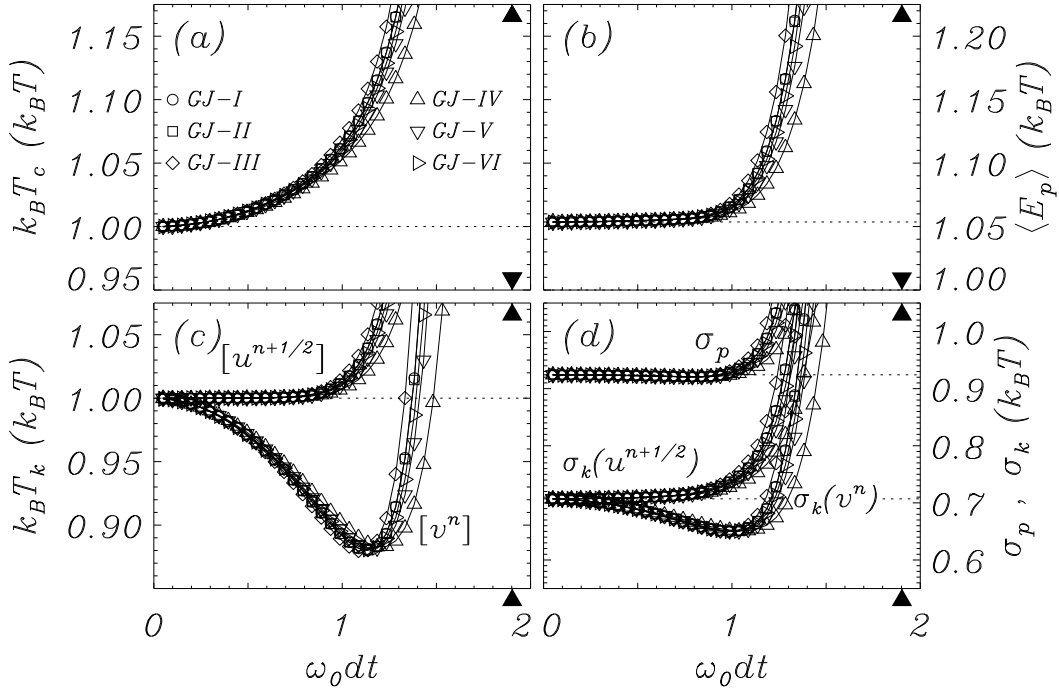}}
%\vspace{-0.50 cm}
\caption{GJ simulations of the one-dimensional oscillator described by $E_{p2}(r)$ in Eq.~(\ref{eq:E_p2}) with $\alpha/m\omega_0=0.1$, $\theta=\frac{3}{4}\pi$, $\kappa r_0^2/E_0=1/40$, and $k_BT=E_0$. Statistics for each data point is acquired from simulations of 10$^{10}$ time steps. Data shown for methods 
highlighted in Sec.~\ref{sec:sv_highlight} with on-site velocity $v^n$ and half-step velocity $u^{n+\frac{1}{2}}$ given in Eqs.~(\ref{eq:GJ_sv_v}) and (\ref{eq:GJ_u}). The entire low damping stability range in $\omega_0dt$ is shown for the simulated temperature. See also Figs.~\ref{fig_1} and \ref{fig_2} for method coefficients and linear stability. Solid pointers indicate the linear stability limit for GJ-III given by Eq.~(\ref{eq:Stability_III}). Actual equations simulated are in the form of Eq.~(\ref{eq:compact_all}). (a) Configurational temperature $T_c$ given by Eq.~(\ref{eq:Tc}). Horizontal dotted line is $k_BT_c=k_BT$. (b) Average potential energy given by the sampling of $E_{p2}$ in Eq.~(\ref{eq:E_p2}). Horizontal dotted line is given by the expected value from Eq.~(\ref{eq:moments}). (c) Kinetic temperature $T_k$ given by Eq.~(\ref{eq:Tk}). Kinetic temperatures are given using both on-site and half-step velocities. Horizontal dotted line is $k_BT_k=k_BT$. (d) Fluctuations, $\sigma_k$ and $\sigma_p$, of kinetic and potential energies. Kinetic fluctuations are shown for both on-site and half-step velocities. Horizontal dotted lines for $\sigma_k$ represent the expectation $\sigma_k=k_BT/\sqrt{2}$. Horizontal dotted line for $\sigma_p$ represents the expected result given by Eq.~(\ref{eq:moments}).
}
\label{fig_7}
\end{figure}

\begin{figure}[t]
\centering
\scalebox{0.8}{\centering \includegraphics[trim={1.5cm 7.0cm 1.5cm 2.0cm},clip]{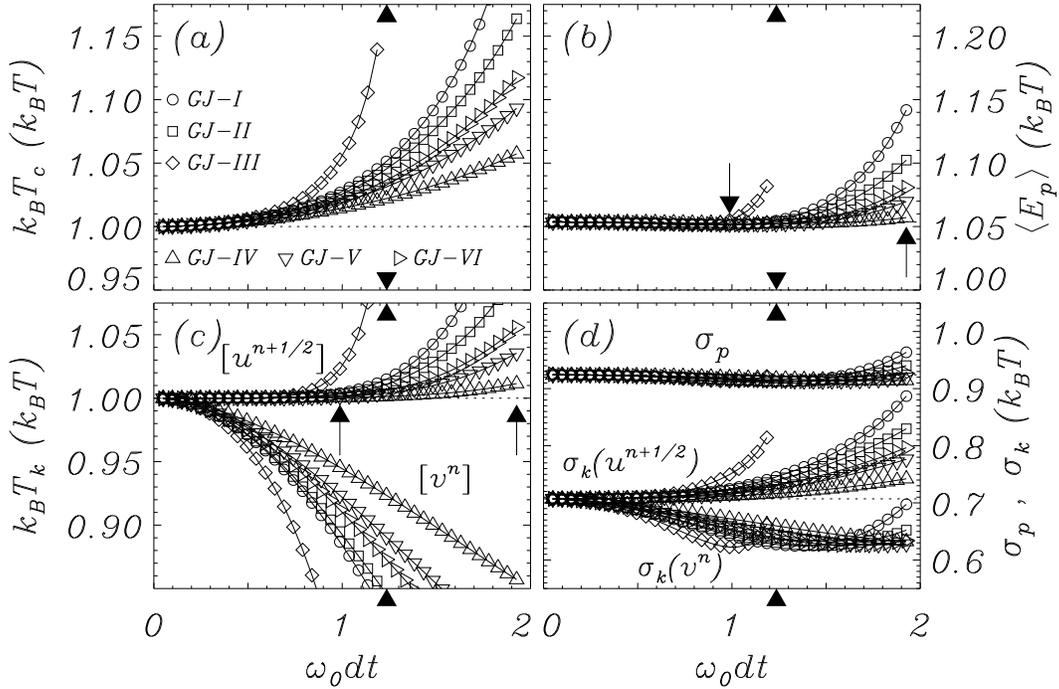}}
%\vspace{-0.50 cm}
\caption{GJ simulations of the one-dimensional oscillator described by $E_{p2}(r)$ in Eq.~(\ref{eq:E_p2}) with $\alpha/m\omega_0=1$, $\theta=\frac{3}{4}\pi$, $\kappa r_0^2/E_0=1/40$, and $k_BT=E_0$.
Other simulation and figure information are given in the caption of Fig.~\ref{fig_7}. 
Vertical arrows in (b) and (c) refer to the parameters for which the effective potentials are shown in Figs.~\ref{fig_10} and \ref{fig_11} for method GJ-II.
}
\label{fig_8}
\end{figure}

\begin{figure}[t]
\centering
\scalebox{0.8}{\centering \includegraphics[trim={1.5cm 7.0cm 1.5cm 2.0cm},clip]{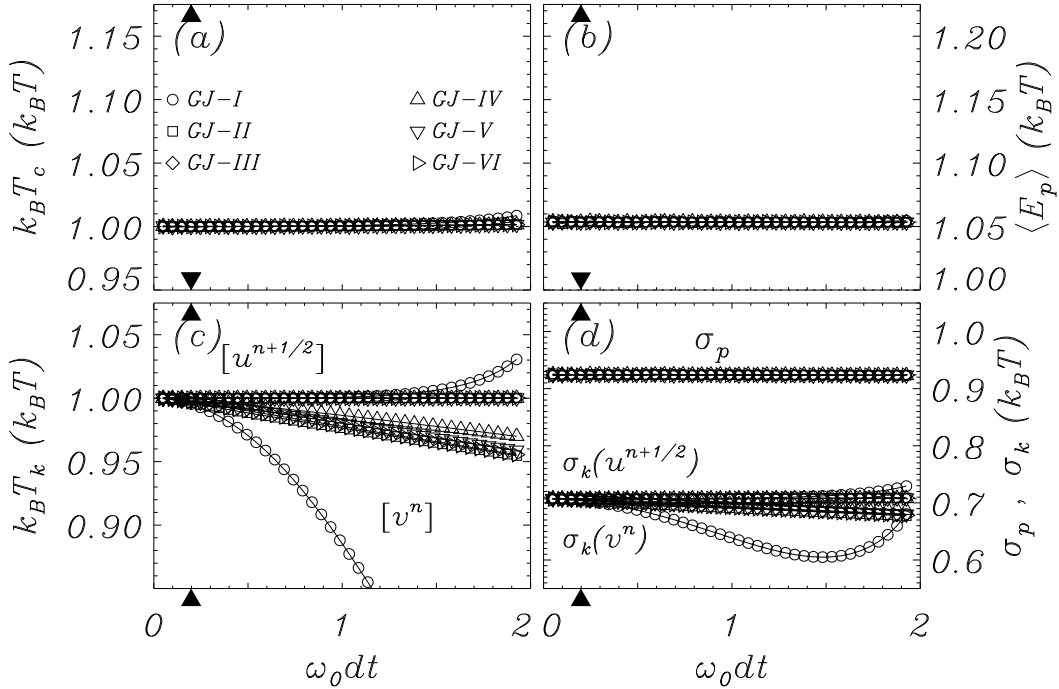}}
%\vspace{-0.50 cm}
\caption{GJ simulations of the one-dimensional oscillator described by $E_{p2}(r)$ in Eq.~(\ref{eq:E_p2}) with $\alpha/m\omega_0=10$, $\theta=\frac{3}{4}\pi$, $\kappa r_0^2/E_0=1/40$, and $k_BT=E_0$.
Other simulation and figure information are given in the caption of Fig.~\ref{fig_7}.
}
\label{fig_9}
\end{figure}

\begin{figure}[t]
\centering
\scalebox{0.8}{\centering \includegraphics[trim={1.5cm 4.0cm 1.5cm 5.0cm},clip]{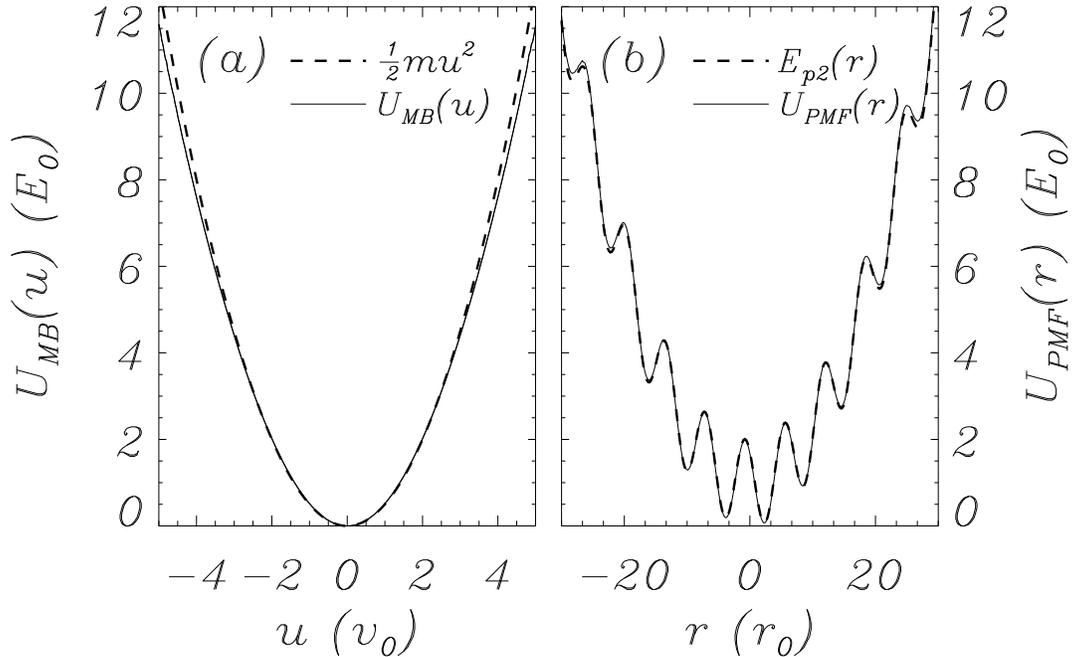}}
%\vspace{-0.50 cm}
\caption{GJ simulation of effective potentials $U_{MB}(u)$ and $U_{PMF}(r)$ in configuration $r^n$ and velocity $u^{n+\frac{1}{2}}$ using method GJ-II, highlighted in Sec.~\ref{sec:sv_highlight}, for $E_{p2}(r)$ in Eq.~(\ref{eq:E_p2}). Parameters are given in the caption for Fig.~\ref{fig_8} with $\omega_0dt=1$, which is approximately half the stability limit, as indicated by arrows in Fig.~\ref{fig_8}bc.
(a) Effective kinetic potential as a function of the half-step velocity. Solid curve is $U_{MB}(u)$ as measured from the distribution function $\rho_k$ given in Eq.~(\ref{eq:U_MB}); dashed curve is the continuous time expectation. (b) Effective potentials of mean force as a function of the coordinate $r^n$. Solid curve is $U_{PMF}(r)$ as measured from the distribution function $\rho_c$ given in Eq.~(\ref{eq:U_PMF}); dashed curve is the continuous time expectation from Eq.~(\ref{eq:E_p2}).
}
\label{fig_10}
\end{figure}

\begin{figure}[t]
\centering
\scalebox{0.8}{\centering \includegraphics[trim={1.5cm 4.0cm 1.5cm 5.0cm},clip]{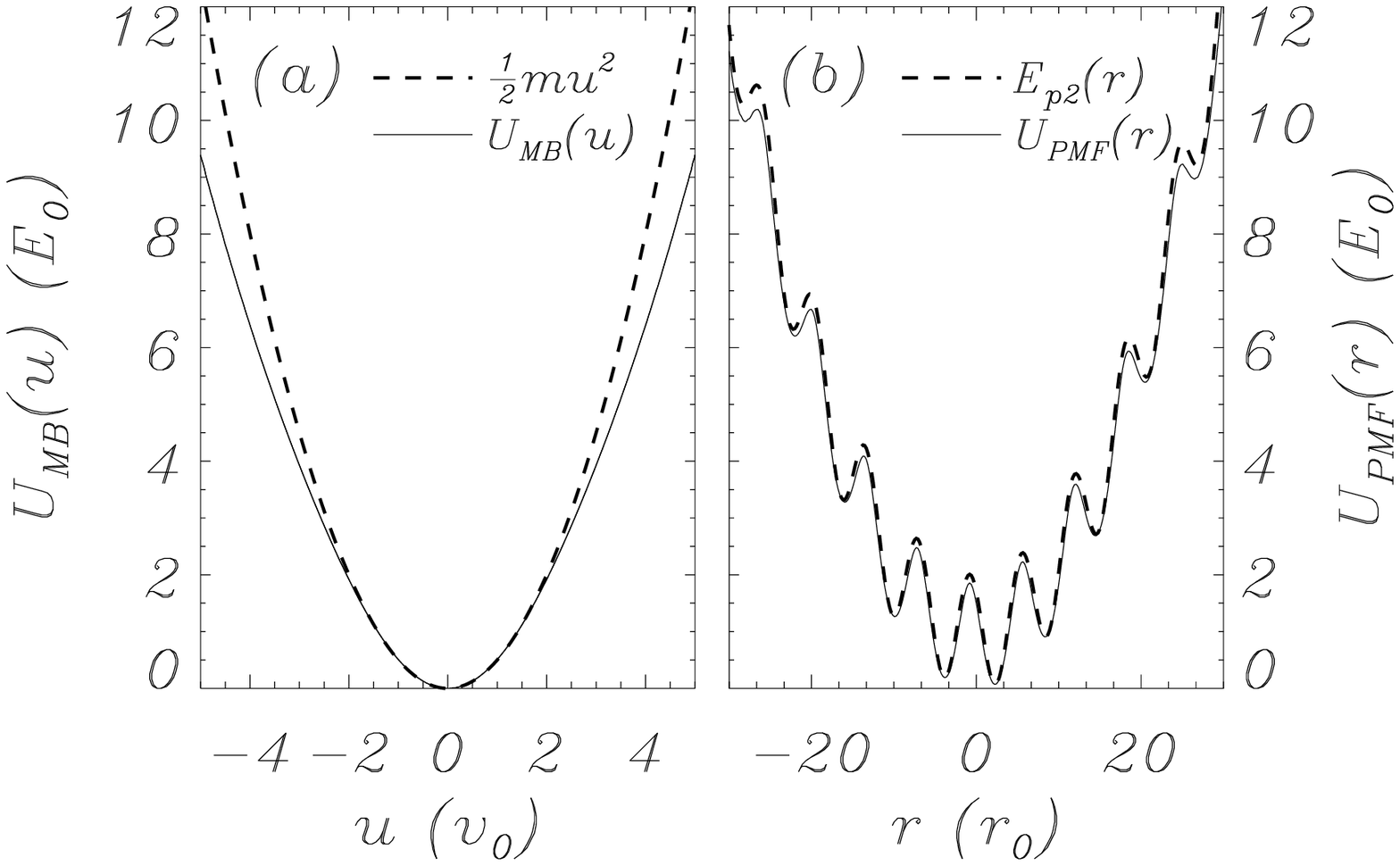}}
%\vspace{-0.50 cm}
\caption{GJ simulation of effective potentials $U_{MB}(u)$ and $U_{PMF}(r)$ in configuration $r^n$ and velocity $u^{n+\frac{1}{2}}$ using GJ method ${\rm II}$, highlighted in Sec.~\ref{sec:sv_highlight}, for $E_{p2}(r)$ in Eq.~(\ref{eq:E_p2}). Parameters are given in the caption for Fig.~\ref{fig_8} with $\omega_0dt=1.95$, which is at the edge of the stability limit, as indicated by arrows in Fig.~\ref{fig_8}bc. Other information is given in the caption of Fig.~\ref{fig_10}.
}
\label{fig_11}
\end{figure}

All the simulation results shown in this section are done for $k_BT=E_0$, and statistical averages are acquired for simulations of 10$^{10}$ time steps. The statistical measures are, respectively, the kinetic and configurational temperatures
\begin{eqnarray}
T_k & = & \frac{2}{k_B}\langle E_k\rangle \label{eq:Tk} \\
T_c & = & \frac{1}{k_B}\frac{\langle(\frac{\partial E_p}{\partial r})^2\rangle}{\langle\frac{\partial^2E_p}{\partial r^2}\rangle} \; , \label{eq:Tc}
\end{eqnarray}
$T_k$ being the kinetic temperature calculated from a given velocity definition, and $T_c$ being the configurational temperature \cite{Hirschfelder,Rickayzen}. The corresponding kinetic and potential fluctuations $\sigma_k$ and $\sigma_p$ are given by
\begin{eqnarray}
\sigma_k^2 & = & \langle E_k^2\rangle-\langle E_k\rangle^2\\
\sigma_p^2 & = & \langle E_p^2\rangle-\langle E_p\rangle^2 \; .\label{eq:sigma_p}
\end{eqnarray}
Additionally, we measure the velocity and coordinate density distribution functions, $\rho_k$ and $\rho_c$, in order to evaluate if the phase space is sampled in accordance with Maxwell-Boltzmann and Boltzmann distributions, which read
\begin{eqnarray}
\rho_k(u) & = & \sqrt{\frac{m}{2\pi k_BT}}\,\exp(-\frac{mu^2}{2k_BT}) \\
\rho_c(r) & = & \frac{\displaystyle\exp(-\frac{E_p(r)}{k_BT})}{\displaystyle\int_{-\infty}^{\infty}\exp(-\frac{E_p(r)}{k_BT})\,dr}\; . \label{eq:dist_c}
\end{eqnarray}
Thus, from the measured density distribution functions, we form the two potentials
\begin{eqnarray}
U_{MB}(u) & = & -k_BT\ln\rho_k + {\cal C}_k \label{eq:U_MB}\\
U_{PMF}(r) & = & -k_BT\ln\rho_c + {\cal C}_c \; , \label{eq:U_PMF}
\end{eqnarray}
where the two constants, ${\cal C}_k$ and ${\cal C}_c$, are determined such that ${\rm min}[U_{PMF}]={\rm min}[E_p]$ and $U_{MB}(0) = E_k(0)=0$. Clearly, a successful simulation will result in $U_{MB}(u)=\frac{1}{2}mu^2$ and $U_{PMF}(r)=E_p(r)$. The physical, continuous-time values for the fluctuations in kinetic and potential energies are expected to be
\begin{eqnarray}
\sigma_k & = & \frac{1}{\sqrt{2}}\,k_BT 
\end{eqnarray}
for the kinetic energy, and given by numerically evaluated moments
\begin{eqnarray}
\langle E_p^k(r)\rangle & = & \frac{\displaystyle\int_{-\infty}^\infty E_p^k(r)\exp(-\frac{E_p(r)}{k_BT})\,dr}{\displaystyle\int_{-\infty}^\infty\exp(-\frac{E_p(r)}{k_BT})\,dr} \label{eq:moments}
\end{eqnarray}
to be inserted into Eq.~(\ref{eq:sigma_p}).

Figures~\ref{fig_3}-\ref{fig_5} show the acquired statistics for the system defined by the potential $E_{p1}(r)$, given in Eq.~(\ref{eq:E_p1}). The three figures show configurational temperatures, average potential energies, kinetic temperatures, and energy fluctuations as a function of the reduced time step $\omega_0dt$ over the entire range of stability. The simulations have been carried out for friction parameters $\alpha/m\omega_0=0.1$ (Fig.~\ref{fig_3}), $\alpha/m\omega_0=1$ (Fig.~\ref{fig_4}), and $\alpha/m\omega_0=10$ (Fig.~\ref{fig_5}). It is noticeable that this stability range is considerably smaller than what linear analysis suggests. While linear analysis predicts $\Omega_0dt=\sqrt{72}\,\omega_0dt<2$ $\Rightarrow$ $\omega_0dt<0.2357$, the actual stability range seems to be limited by $\omega_0dt\le0.02$. The reason is that the potential has a divergent convex nonlinearity, which changes the local curvature (and therefore the stability) as the temperature is increased. Thus, at this temperature $k_BT=E_0$, the stability range is considerably reduced from the linear result. Along with the simulation results, we indicate with horizontal dotted lines the continuous-time expectations for reference. In Figs.~\ref{fig_3}a-\ref{fig_5}a, the horizontal dotted line refers to the expected temperature $T_c=T$, in Figs.~\ref{fig_3}b-\ref{fig_5}b, the horizontal dotted line refers to the expected average potential energy given by Eq.~(\ref{eq:moments}) for $k=1$ (in this case, we find $\langle E_{p1}\rangle\approx0.3424340E_0$), in Figs.~\ref{fig_3}c-\ref{fig_5}c, the horizontal dotted line refers to the expected temperature $T_k=T$, and in Figs.~\ref{fig_3}d-\ref{fig_5}d, the horizontal dotted lines refer to the expected fluctuations in kinetic energy $\sigma_k=k_BT/\sqrt{2}$, and fluctuations in potential energy given by Eq.~(\ref{eq:moments}) for $k=1,2$ and Eq.~(\ref{eq:sigma_p}) (in this case, we find $\sigma_{p1}\approx0.5489407E_0$). All six methods yield extremely accurate and nearly indistinguishable statistics in the entire range of stability for the chosen damping parameters, and the expected deviations for kinetic results using the on-site velocity are observed; see Figs.~\ref{fig_3}cd-\ref{fig_5}cd. The largest damping value, shown in Fig.~\ref{fig_5}, exhibits the increased stability for methods GJ-II, IV-VI that is expected from Eq.~(\ref{eq:Stability}) and Fig.~\ref{fig_2}. However, this increase is not necessarily an expression of more efficient sampling, since the larger damping also slows down the dynamics. In order to demonstrate the details of the statistics in the simulations, we include Fig.~\ref{fig_6}, which displays with solid curves the effective potentials $U_{MB}(u)$ (Fig.~\ref{fig_6}a) and $U_{PMF}(r)$ (Fig.~\ref{fig_6}b) as derived from the distribution functions using method ${\rm III}$; see Eqs.~(\ref{eq:U_MB}) and (\ref{eq:U_PMF}). The parameter values are as in Fig.~\ref{fig_3}, and the data point for which these data are shown is indicated with vertical arrows in Fig.~\ref{fig_3}bc. Along with the simulated effective potentials, we also show the continuous-time expectations with dashed curves. It is obvious that, even at the edge of the stability limit, the sampled distributions $\rho_k(u)$ and $\rho_c(r)$ are in near-perfect agreement with the correct results. The results shown in Fig.~\ref{fig_6} are representative of all cases we have simulated for the potential $E_{p1}$.

Figures~\ref{fig_7}-\ref{fig_9} show the acquired statistics for the system defined by the potential $E_{p2}(r)$, given in Eq.~(\ref{eq:E_p2}). The three figures show configurational temperature, average potential energies, kinetic temperatures, and energy fluctuations as a function of the reduced time step $\omega_0dt$ over the entire range of standard Verlet stability $\Omega_0dt<2$. The simulations have been carried out for the parameter $\kappa r_0^2/E_0=1/40$, and friction values $\alpha/m\omega_0=0.1$ (Fig.~\ref{fig_7}), $\alpha/m\omega_0=1$ (Fig.~\ref{fig_8}), and $\alpha/m\omega_0=10$ (Fig.~\ref{fig_9}). Unlike the potential $E_{p1}$, $E_{p2}$ has a periodic (soft) nonlinearity, which means that the linear stability range in time step can be expected to not be strongly dependent on the applied temperature. Along with the simulation results, we indicate with horizontal dotted lines the continuous-time expectations for reference. In Figs.~\ref{fig_7}a-\ref{fig_9}a, the horizontal dotted line refers to the expected temperature $T_c=T$; in Figs.~\ref{fig_7}b-\ref{fig_9}b, the horizontal dotted line refers to the expected average potential energy given by Eq.~(\ref{eq:moments}) for $k=1$ (in this case, we find $\langle E_{p2}\rangle\approx1.053610E_0$); in Figs.~\ref{fig_7}c-\ref{fig_9}c, the horizontal dotted lines refer to the expected temperature $T_k=T$; and in Figs.~\ref{fig_7}d-\ref{fig_9}d, the horizontal dotted lines refer to the expected fluctuations in kinetic energy $\sigma_k=k_BT/\sqrt{2}$, and fluctuations in potential energy given by Eq.~(\ref{eq:moments}) for $k=1,2$ and Eq.~(\ref{eq:sigma_p}) (in this case, we find $\sigma_{p2}\approx0.9243081E_0$). Markers on the horizontal axes point to the linear stability limit for method ${\rm III}$, as given by Eq.~(\ref{eq:Stability_III}). For low damping the six methods behave almost the same with clear advantages of the half-step velocity over the on-site velocity. However, for reduced time steps above about half of the stability limit, significant deviations from the continuous-time expectations arise abruptly. This has previously been observed for the GJF trajectory in Ref.~\cite{GJF4}, and the general discrete-time phenomenon has recently been analyzed in detail in Ref.~\cite{2GJ2}. The origin of the discrepancies is due to nonlinear, discrete-time resonant modes that can be dynamically stabilized for appreciable time steps despite a small amount of damping. These nonlinear, discrete-time resonances are significantly attenuated as friction is increased, as can be observed in Figs.~\ref{fig_8} and \ref{fig_9}. The phenomenon is an inherent consequence of nonlinearity and discrete-time, is common to discrete-time algorithms, and falls outside of linear stability analysis. Despite this, the overall results are in agreement with the attractive features of this class of methods. As predicted, method ${\rm III}$ suffers in stability when the friction parameter becomes large. In Fig.~\ref{fig_9} this method is stable only for a very small part of the stability ranges for methods ${\rm I}$ and ${\rm II}$. We also observe that the accuracy of the on-site velocity is better for methods ${\rm II}$, IV-VI than it is for method ${\rm I}$ (especially for high friction), as predicted by Eq.~(\ref{eq:Ek_GJ}). In Fig.~\ref{fig_8}bc we point with vertical arrows to selected data points for which we show the detailed statistics in Figs.~\ref{fig_10} and \ref{fig_11}. Similarly to Fig.~\ref{fig_6}, we display with solid curves the effective potentials $U_{MB}(u)$ (Figs.~\ref{fig_10}a and \ref{fig_11}a) and $U_{PMF}(r)$ (Figs.~\ref{fig_10}b and \ref{fig_11}b) as derived from the distribution functions using method ${\rm II}$; see Eqs.~(\ref{eq:U_MB}) and (\ref{eq:U_PMF}). The parameter values are as in Fig.~\ref{fig_8}. Along with the simulated effective potentials, we also show the continuous-time expectations with dashed curves. For a reduced time step well within the stability limit, as shown in Fig.~\ref{fig_10} for $\omega_0dt=1$, the comparisons between the observed distributions in both half-step velocity and position with their continuous-time expectations are remarkably good. Even in the large time step regime, shown in Fig.~\ref{fig_11} for $\omega_0dt=1.95$, we observe relatively minor discrete-time errors in the effective potentials, consistent with the averages shown in Fig.~\ref{fig_8}. The observed discrepancies in Fig.~\ref{fig_11} are related to the nonlinear, discrete-time instabilities discussed in Ref.~\cite{2GJ2}, and Fig.~\ref{fig_9} confirms that simulations with large friction do not display this complication.

%%%%%%%%%%%%%%%%%%%%%%%%%%%%%%%%%%%%%%%%%%%%%%%%%%%%
\subsection{Molecular Dynamics model system}
\label{sec:molecular-dynamics}

\begin{figure}[t]
\centering
\scalebox{0.6}{\centering \includegraphics[trim={1.5cm 5.5cm 1.5cm 2.5cm},clip]{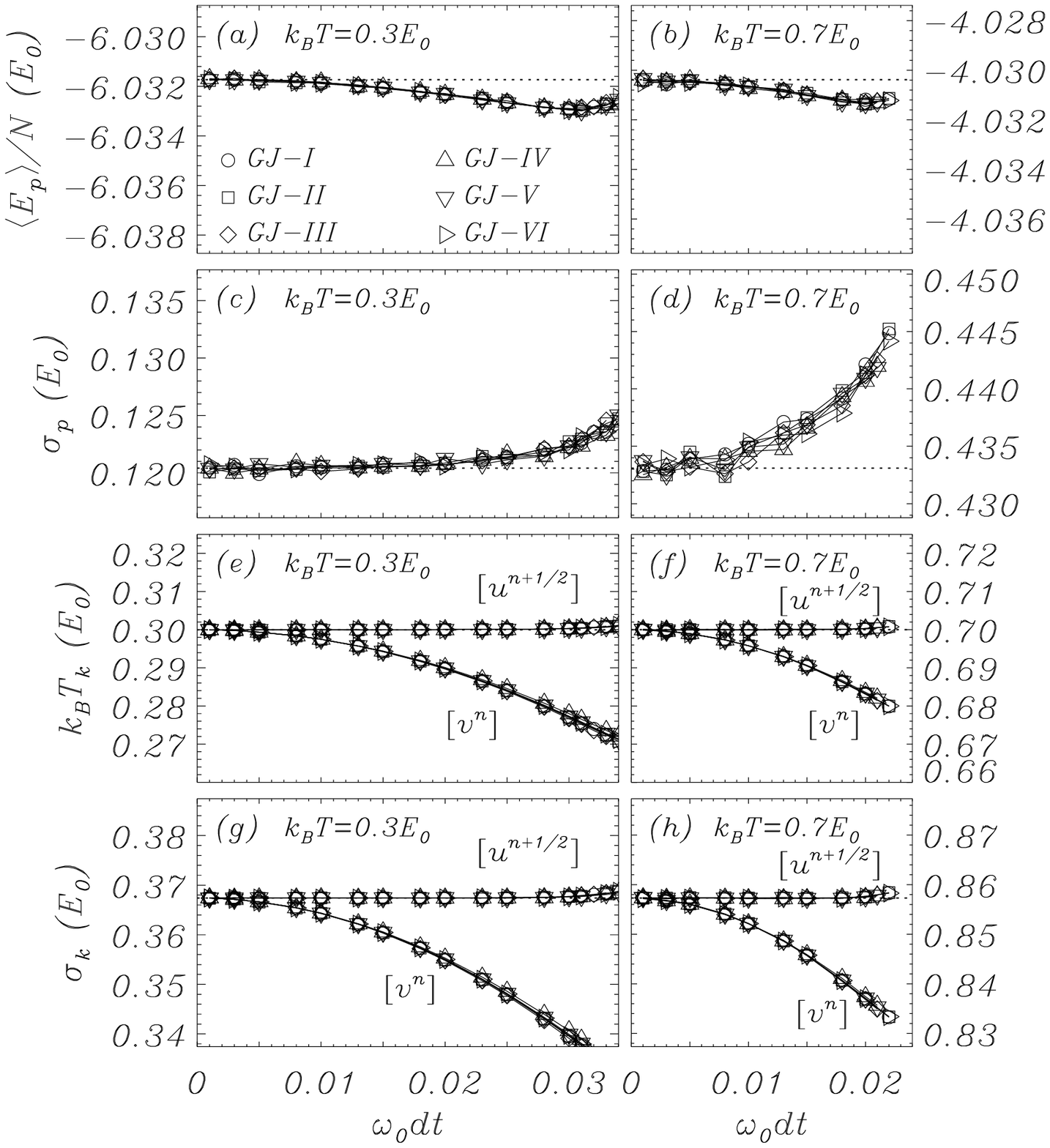}}
%\vspace{-0.50 cm}
\caption{GJ simulations for methods I-VI applied to Molecular Dynamics. Statistical averages of potential energy $\langle E_p\rangle$ (a) and (b), and its standard deviation $\sigma_p$ (c) and (d), kinetic temperature $k_BT_k$ (e) and (f), and kinetic energy fluctuations of each particle $\sigma_k$ (g) and (h) as a function of reduced time step $\omega_0dt$ for $\alpha=1\,m\omega_0$, sampled over $\omega_0\Delta t=2\times10^{5}$ units of time. $N=864$ particles are simulated with interaction potential Eq.~(\ref{eq:Eq_LJ_spline}) in a fixed cubic box with periodic boundary conditions. (a,c,e,g) show results for a crystalline FCC state at $k_BT=0.3E_0$ and volume $V=617.2558r_0^3$; (b,d,f,h) show results for a liquid state at $k_BT=0.7E_0$ and volume $V=824.9801r_0^3$. Results are shown for GJ methods I-VI, highlighted in Sec.~\ref{sec:sv_highlight} with on-site velocity $v^n$ and half-step velocity $u^{n+\frac{1}{2}}$ given in Eqs.~(\ref{eq:GJ_sv_v}) and (\ref{eq:GJ_u}). Horizontal dotted lines in (a-d) indicate the results for small $\omega_0dt$. Horizontal dotted lines in (e-h) indicate the known results $k_BT_k=k_BT$ and $\sigma_k = k_BT\sqrt{3}/\sqrt{2}$. The entire light damping stability range in $\omega_0dt$ is shown for the simulated temperatures. Simulated equations are in the form of Eq.~(\ref{eq:compact_all}).
}
\label{fig_12}
\end{figure}

\begin{figure}[t]
\centering
\scalebox{0.6}{\centering \includegraphics[trim={1.5cm 14.5cm 1.5cm 2.5cm},clip]{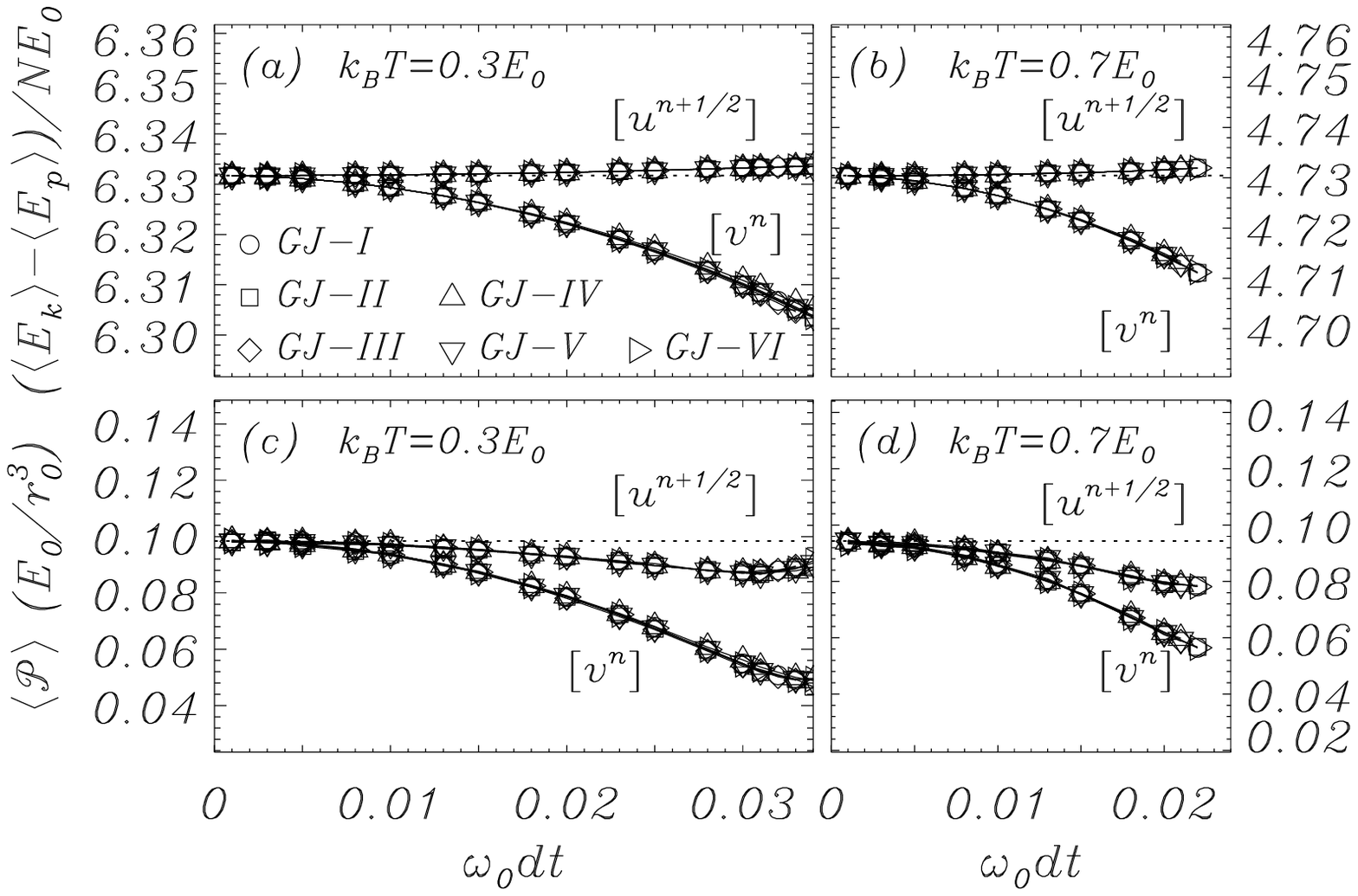}}
%\vspace{-0.50 cm}
\caption{GJ simulations for methods I-VI applied to Molecular Dynamics. Statistical averages of the virial $\langle E_k\rangle-\langle E_p\rangle$ (a,b) and the pressure ${\cal P}$ (c,d) given in Eq.~(\ref{eq:Press}) as a function of reduced time step for the simulations shown in Fig.~\ref{fig_12}. The kinetic component of the mixed quantities are calculated both with the on-site and half-step velocities, as indicated.
}
\label{fig_13}
\end{figure}

\begin{figure}[t]
\centering
\scalebox{0.6}{\centering \includegraphics[trim={1.5cm 5.5cm 1.5cm 2.5cm},clip]{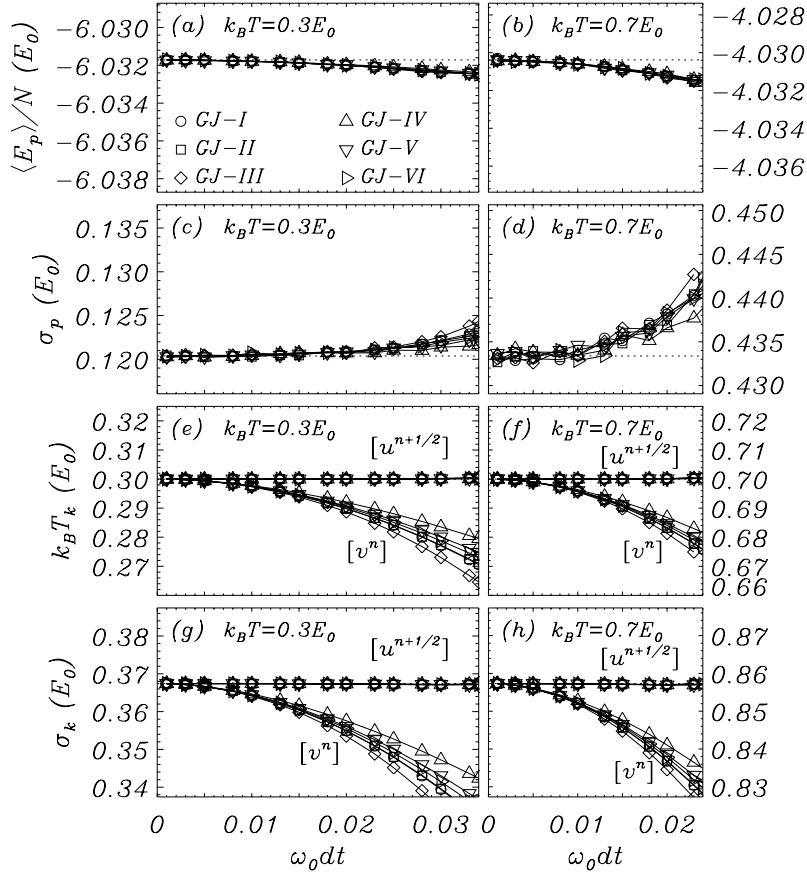}}
%\vspace{-0.50 cm}
\caption{GJ simulations for methods I-VI applied to Molecular Dynamics with $\alpha=10\,m\omega_0$.
Other system parameters are as in Fig.~\ref{fig_12}.
}
\label{fig_14}
\end{figure}

\begin{figure}[t]
\centering
\scalebox{0.6}{\centering \includegraphics[trim={1.5cm 14.5cm 1.5cm 2.5cm},clip]{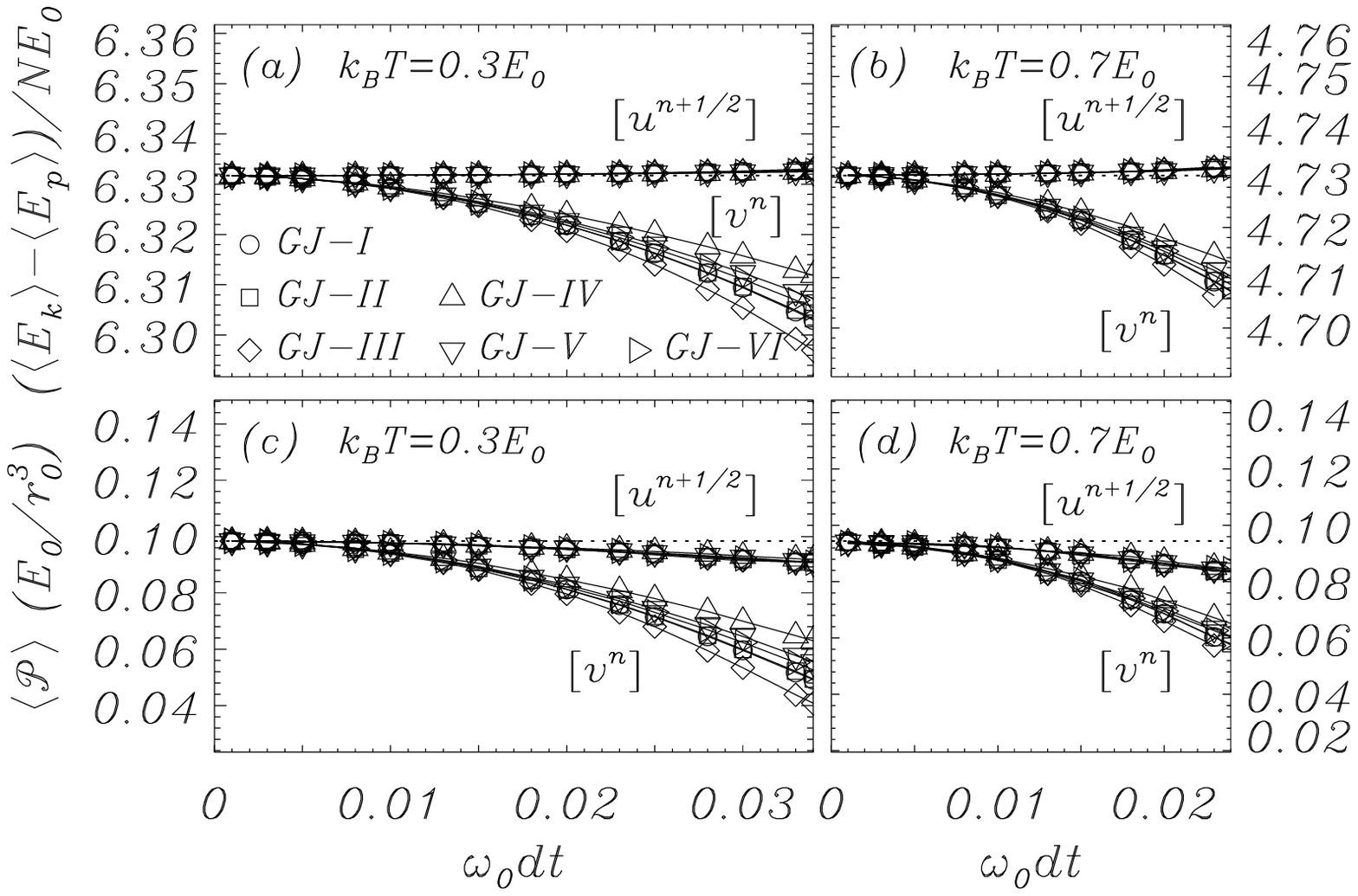}}
%\vspace{-0.50 cm}
\caption{GJ simulations for methods I-VI applied to Molecular Dynamics. Statistical averages of the virial $\langle E_k\rangle-\langle E_p\rangle$ (a,b) and the pressure ${\cal P}$ (c,d) given in Eq.~(\ref{eq:Press}) as a function of reduced time step for the simulations shown in Fig.~\ref{fig_14}. The kinetic component of the mixed quantities are calculated both with the on-site and half-step velocities, as indicated.
}
\label{fig_15}
\end{figure}

\begin{figure}[t]
\centering
\scalebox{0.6}{\centering \includegraphics[trim={1.5cm 5.5cm 1.5cm 2.5cm},clip]{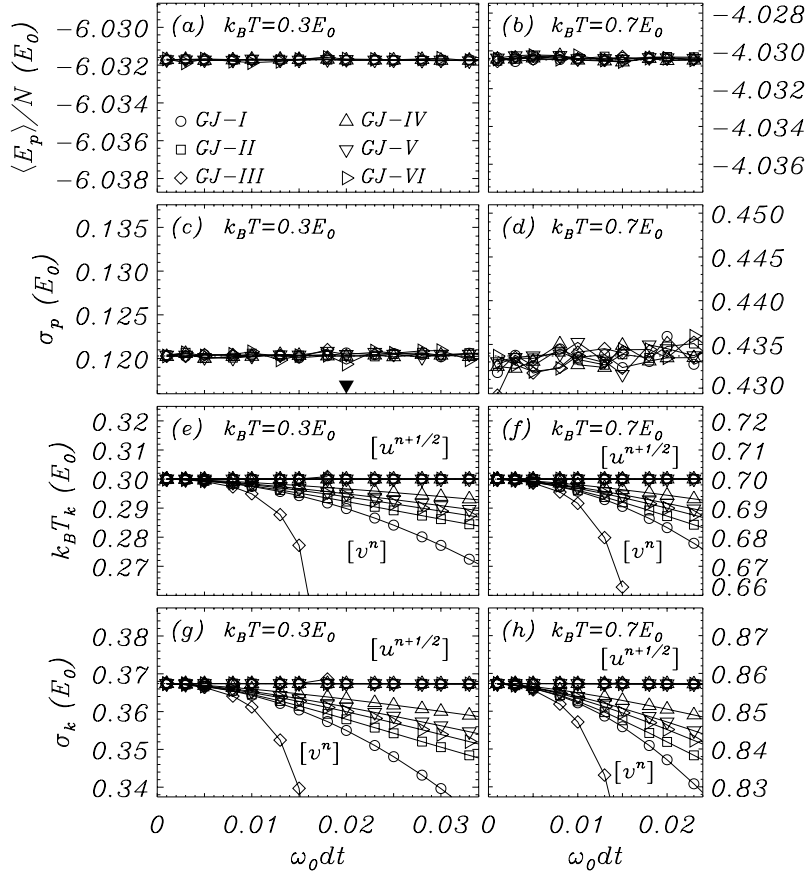}}
%\vspace{-0.50 cm}
\caption{GJ simulations for methods I-VI applied to Molecular Dynamics with $\alpha=100\,m\omega_0$.
Other system parameters are as in Fig.~\ref{fig_12}. Solid triangular marker in (c) shows the stability limit for GJ-III as given by Eq.~(\ref{eq:Stability_III}).
}
\label{fig_16}
\end{figure}

\begin{figure}[t]
\centering
\scalebox{0.6}{\centering \includegraphics[trim={1.5cm 14.5cm 1.5cm 2.5cm},clip]{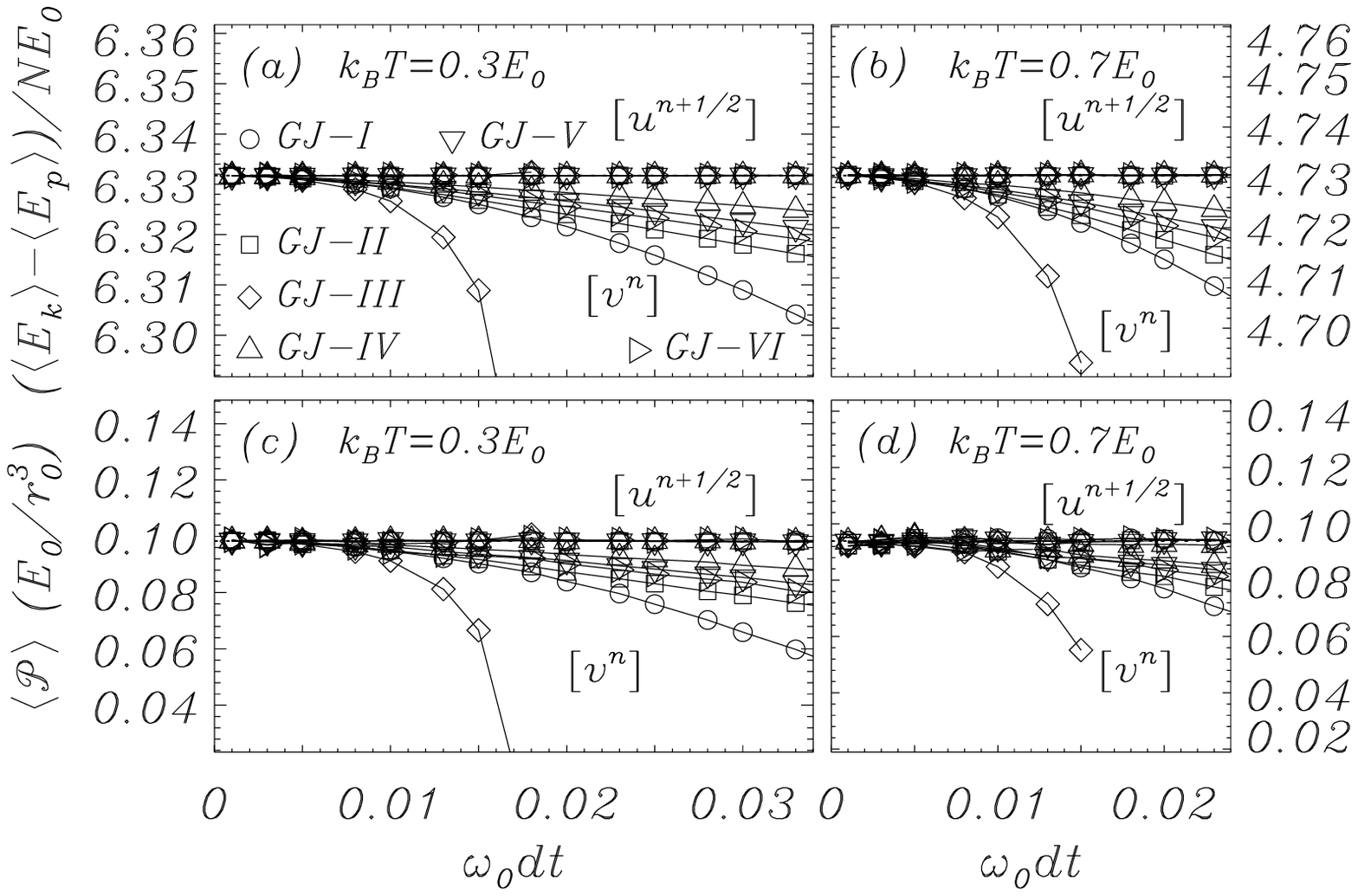}}
%\vspace{-0.50 cm}
\caption{GJ simulations for methods I-VI applied to Molecular Dynamics. Statistical averages of the virial $\langle E_k\rangle-\langle E_p\rangle$ (a,b) and the pressure ${\cal P}$ (c,d) given in Eq.~(\ref{eq:Press}) as a function of reduced time step for the simulations shown in Fig.~\ref{fig_16}. The kinetic component of the mixed quantities are calculated both with the on-site and half-step velocities, as indicated.
}
\label{fig_17}
\end{figure}

In order to further demonstrate the applicability of the GJ set of methods for Molecular Dynamics, we have conducted simulations for all the exemplified GJ methods I-VI. These simulations are conducted similarly to those shown in Ref.~\cite{2GJ}, both for consistency and  for comparison with other classic thermostats \cite{BBK,Pastor_88}. Results are shown for both solid and liquid states of a simple, one-component system in the $(N,V,T)$ ensemble. We adopt the splined, short range Lennard-Jones potential described in Ref.~\cite{GJF3}:
\begin{eqnarray}
\frac{E_p(r)}{E_0} & = &
\left\{\begin{array}{lcl} \displaystyle{\left(\frac{r_0}{|r|}\right)^{12}-2\left(\frac{r_0}{|r|}\right)^{6}} & , & 0<|r|\le r_s\\ 
\\ \displaystyle{\frac{a_4}{E_0}\left(|r|-r_c\right)^4+\frac{a_8}{E_0}\left(|r|-r_c\right)^8} & , & r_s<|r|<r_c \\ \\ \displaystyle{0} & , & r_c \le |r|
\end{array}\right. \; , \label{eq:Eq_LJ_spline}
\end{eqnarray}
where $r$ is a three-dimensional coordinate between any two particles. The parameters are given by
\begin{subequations}
\begin{eqnarray}
\frac{r_s}{r_0} & = & \left(\frac{13}{7}\right)^{1/6} \; \approx \; 1.108683 \\
\frac{r_c}{r_0} & = & \frac{r_s}{r_0}-\frac{32E_p(r_s)}{11E_p^\prime(r_s)r_0} \; \approx \; 1.959794\\
a_4 & = & \frac{8E_p(r_s)+(r_c-r_s)E_p^\prime(r_s)}{4(r_c-r_s)^4} \\
a_8 & = & -\frac{4E_p(r_s)+(r_c-r_s)E_p^\prime(r_s)}{4(r_c-r_s)^8} \; . 
\end{eqnarray}
\end{subequations}
The potential $E_p(r)$ has minimum $-E_0$ at $|r|=r_0$, and is smoothly splined
between the inflection point of the Lennard-Jones potential and zero with continuity through the second derivative
at $|r|=r_s$, and continuity through the third derivative at $|r|=r_c$. Simulations are done with $N=864$ particles of identical mass $m$ in a cubic box of side-length $L$, and with periodic boundary conditions. Initial conditions are chosen to be the Face Centered Cubic (FCC) structure, which is allowed to evolve for a long transient time before acquiring statistical data in order to reach a thermodynamically representative ensemble for a given applied temperature and a reduced time step $\omega_0dt$, where $\omega_0^2mr_0^2=E_0$.

Throughout the stability ranges for the time step, we conduct simulations for two characteristically different temperatures, representing a crystalline solid at a set volume of $V=L^3=(8.51442r_0)^3$ for $k_BT=0.3E_0$, and a liquid at a set volume of $V=L^3=(9.378812r_0)^3$ for $k_BT=0.7E_0$. Each simulation makes statistical averages over a time span of $\omega_0\Delta t=2\times10^5$.

For very low damping, $\alpha/m\omega_0=1$, we observe in Figs.~\ref{fig_12} and \ref{fig_13} that all six methods yield near identical results in all displayed quantities for both solid ($k_BT=0.3E_0$) and liquid ($k_BT=0.7E_0$) systems. As we have previously observed for the GJF method (GJ-I), the other methods are seen to show the same slight decrease in potential energy average as a function of increasing time step, Fig.~\ref{fig_12}ab. This is accompanied by a slight increase in the associated fluctuations seen in Fig.~\ref{fig_12}cd. The kinetic properties shown in Fig.~\ref{fig_12}e-h display the remarkable accuracy of the half-step velocity for all the simulated cases. This is true for both average kinetic energy (Fig.~\ref{fig_12}ef) and the fluctuations of the kinetic energy of each particle (Fig.~\ref{fig_12}gh). We notice that the fluctuations $\sigma_k$ in kinetic energy of each three-dimensional particle are expected to be
\begin{eqnarray}
\sigma_k & = & \frac{\sqrt{3}}{\sqrt{2}}\,k_BT\; ,
\end{eqnarray}
which is shown as a horizontal dotted line in Figs.~\ref{fig_12}gh, \ref{fig_13}gh, and \ref{fig_16}gh. The simulation data are in stunning agreement with the expected result. The kinetic properties are also shown in the same figures for the on-site velocity of the six methods. These results show the expected deviations consistent with our observation that any on-site velocity will have a second order (in $\omega_0dt$) error.

Figure \ref{fig_13} investigates the relationships between the kinetic and configurational statistics with (a) and (b) exploring the difference $\langle E_k\rangle-\langle E_p\rangle$, and (c) and (d) showing the system pressure ${\cal P}$ calculated from
\begin{eqnarray}
{\cal P} & = & \frac{1}{3V}\left\langle\sum_{i=1}^Nf_i\cdot r_i\right\rangle+\frac{Nk_BT_k}{V} \; , \label{eq:Press}
\end{eqnarray}
where $f_i$ is the total force on the particle with coordinate $r_i$, and where $T_k$ is calculated from Eq.~(\ref{eq:Tk}).  Again, we observe that the  combination of the GJ trajectories with their half-step velocities are superior over the use of the on-site velocities, since the half-step velocities are successfully designed to be time-step-independent. When using the on-site velocities, the deviations are mostly due to the above-mentioned time step dependence, while the small deviations for the half-step velocities are due to the slight time-step dependence seen for the configurational quantities in Fig.~\ref{fig_12}. 

Increasing the damping parameter to $\alpha/m\omega_0=10$, we show the results in Figs.~\ref{fig_14} and \ref{fig_15}. The results are overall the same as what we observed for $\alpha/m\omega_0=1$, with only very minor dependency on the time step in the configurational and half-step-kinetic quantities. The on-site velocity results show the predicted (from Eq.~(\ref{eq:Ek_GJ})) time step dependency, and that this dependence is different for the different choices of method ($c_2$). From Figs.~\ref{fig_14}e-h and \ref{fig_15} it is obvious that for the on-site velocity, method III is noticeably deviating from the results of the other methods, as expected from Eq.~(\ref{eq:Ek_GJ}).

Finally, for relatively large damping $\alpha/m\omega_0=100$, we show the results in Figs.~\ref{fig_16} and \ref{fig_17}. We here see near-perfect agreement between the continuous-time expectations and simulation results for configurational and kinetic sampling with the half-step velocities. We also observe the considerable differences between the statistics of the six simulation methods. Method III is visibly challenged by the time step in accordance with the stability criterion given in Eq.~(\ref{eq:Stability_III}), and it has for this value of the damping only been possible to conduct simulations with method III for $\omega_0dt\le0.017$ for the crystalline ($k_BT=0.3E_0$) case, and for $\omega_0dt\le0.015$ for the liquid case ($k_BT=0.7E_0$). In contrast, Method I shows the exact same on-site kinetic behavior for all simulated damping values, whereas methods II, IV-VI see improvements of the quality of the on-site kinetic results as damping is increased, in agreement with Eq.~(\ref{eq:Ek_GJ}).

%%%%%%%%%%%%%%%%%%%%%%%%%%%%%%%%%%%%%%%%%%%%%%%%%%%%
\section{Discussion}
\label{sec:discussion}

We have methodically investigated the possibilities for obtaining statistically correct stochastic Verlet-based methods for simulating Langevin systems. Assuming a general five-parameter model for a stochastic St{\o}rmer-Verlet trajectory with a single noise value per time step, we have required that a method must provide correct, time-step-independent results for diffusion on a flat surface and for the average potential energy in a noisy harmonic oscillator. These requirements lead to a set of methods (GJ), where each method can be defined by a single function of the damping, time step, and mass; namely the single time step velocity attenuation parameter $c_2$. 

For each of the derived stochastic St{\o}rmer-Verlet methods in this GJ set of configurationally correct methods, we explored the general five parameter, three-point finite difference approximation to a velocity, consistent with the Verlet framework. As a general finite difference approximation, an expression can result in both on-site and half-step velocities. We determine if a velocity definition is on-site or half-step by evaluating the statistical temporal relationship between the coordinate $r^n$ and an associated velocity. The continuous-time expectation of the cross-correlation between the configurational coordinate and a velocity can be compared to comparable discrete-time expressions, and the symmetry of the expressions is used as a discriminator between different types of velocities, on-site, half-step, or other. Using this definition, we conclude for the GJ methods that there does not exist any finite-difference on-site velocity that is correct and time-step-independent in its statistics. In fact, we conclude that there does not exist any meaningful three-point velocity that can have time-step-independent statistical response. However, we find that there exists a unique on-site velocity for each configurational GJ method such that the kinetic statistics has an error no worse than second order in the reduced time step, and that this is the best possible three-point velocity. In case of the GJF method, this is the on-site velocity that was derived in Ref.~\cite{GJF1}. For each GJ trajectory it is found that there exists a unique half-step velocity that produces time-step-independent and correct statistics. This is a generalization of the GJF-2GJ result recently published in Ref.~\cite{2GJ}, where it was demonstrated that it is possible to obtain robust statistics for both configurational and kinetic sampling in the same simulation, for as long as one uses the identified half step velocity. A noticeable observation from the analysis in this paper is the inherent difficulty of obtaining a half-step (or on-site) velocity that can simultaneously provide correct transport values for diffusion and ballistic (drift) behavior. Only one of the methods (the half-step velocity for GJ-III) contains the intersection of those values with the correct, continuous-time expectations. However, GJ method III becomes unstable as the damping coefficient is increased, and this method is therefore only useful for low to moderate damping. In order to address this complication, we have explored the generalization of the recent work \cite{GJF-2} on defining two-point velocities for the GJF method such that time-step-independent kinetic energy is obtained. By relaxing the condition for statistical symmetry with the configurational coordinate $r^n$, it is possible to obtain a set of kinetically correct two-step velocities in which the deviation from the half-step condition is compensated by an additional noise contribution that ensures the correct kinetic energy. From this set, one can choose the velocity that gives both diffusion and drift correctly in discrete time. However, we emphasize that the temporal location of such velocity is, in general, undetermined, and that using a combination of a coordinate $r^n$ with a velocity that has unknown statistical relationship with the coordinate may corrupt the measure of, e.g., pressure fluctuations and other thermodynamic quantities that mix kinetics with configurational sampling.

The derived set of GJ methods have been written in all familiar and simple forms of the stochastic Verlet-type: SV (Eq.~(\ref{eq:GJ_sv_r}) with associated velocities given in Eqs.~(\ref{eq:GJ_sv_v}) and (\ref{eq:GJ_u})), VV (Eq.~(\ref{eq:GJ_vv})), LF (Eq.~(\ref{eq:GJ_lf})), and compact (Eq.~(\ref{eq:compact_all})). We submit that it should be convenient to implement and validate these methods in any existing simulation code that already has Verlet-type algorithms implemented. We have given a number of reasonable and obvious examples of specific methods, but there may be other choices of the attenuation function $c_2$ that are desirable for certain applications. A representative subset of the methods has been comprehensively validated numerically, and the results compare very well with the expectations from the analysis and previously published work. We submit that the GJ methods derived in this paper constitute the complete set of the most attractive possibilities offered by stochastic Verlet-type algorithms with a single noise term per time step. The methods allow for the most accurate and most efficient acquisition of reliable statistical data at no additional computational cost compared to any other Verlet-type method that we are aware of. The GJF trajectory is currently available within the LAMMPS simulation package as a stochastic thermostat. Including the more general method with its statistically correct half-step velocities is desirable and currently under way.

The development and analysis in this paper have primarily been focused on presenting the possibilities of the statistically most useful methods and their features, and not on giving specific recommendations as to which choice of $c_2$ is the most desirable within this complete set. This choice is likely best left to the requirements of a specific application, but it is important to reemphasize the discussion of the stability criterion for each of the exemplified methods in Sec.~\ref{sec:sv_highlight}, since the sampling efficiencies of the methods may indeed be very similar once time (or frequency) is properly rescaled according to the choice of $c_2$. 
%%%%%%%%%%%%%%%%%%%%%%%%%%%%%%%%%%%%%%%%%%%%%%%%%%%%
\section{Acknowledgment}
The author is grateful for encouraging discussions with Aidan Thompson at the outset of this work, and for conversations with Oded Farago and Richard Scalettar.

\end{document}